\documentclass[structabstract]{aa}

\usepackage{txfonts}
\usepackage{graphicx}
\usepackage{natbib}
\bibpunct{(}{)}{;}{a}{}{,}

\begin{document}

  \title{A close look at the Centaurus A group of galaxies \\ 
IV. Recent star formation histories of late-type dwarfs around CenA}
  \author{D. Crnojevi\'{c}\inst{1,2} \and E. K. Grebel\inst{2} \and A. A. Cole\inst{3}
}

 \institute{Institute for Astronomy, University of Edinburgh, Royal Observatory, Blackford Hill, EH9 3HJ Edinburgh, UK
\newline email: dc@roe.ac.uk
\and
Astronomisches Rechen-Institut, Zentrum f\"{u}r Astronomie der
   Universit\"{a}t Heidelberg, M\"{o}nchhofstrasse 12-14, 69120 Heidelberg, Germany
 \and
School of Mathematics \& Physics, University of Tasmania, Private Bag 37 Hobart, 7001 Tasmania, Australia}

  \date{Received 27 January 2012 / Accepted 16 March 2012}

  \abstract
    {}{We study a sample of five dwarf irregular galaxies in the CenA/M83 group, which are companions to the giant elliptical CenA. We aim at deriving their physical properties over their lifetime and compare them to those of dwarfs located in different environments.}{We use archival HST/ACS data and apply synthetic color-magnitude diagram fitting in order to reconstruct the past star formation activity of the target galaxies.}{The average star formation rate for the studied galaxies ranges from $10^{-3}$M$_\odot$ yr$^{-1}$ up to $\sim7\times10^{-2}$M$_\odot$ yr$^{-1}$, and their mean metallicities correlate with their luminosities (from [Fe/H]$\sim-1.4$ up to $\sim-1.0$). The form of the star formation histories varies across the sample, with quiescent periods alternating with intermittent enhancements in the star formation (from a few up to several times the average lifetime value). The dwarfs in this sample formed $\sim35\%$ to $60\%$ of their stellar content prior to $\sim5$ Gyr ago.}{The resulting star formation histories for the CenA companions are similar to those found for comparable Local Group and M81 group dwarfs. We consider this sample of dwarfs together with five previously studied M83 dwarf irregular companions. We find no trend of the \emph{average} star formation rate with tidal index or distance from the main galaxy of the group. However, dwarfs with higher baryonic masses do show higher \emph{average} star formation rates, underlining the importance of intrinsic properties in governing the evolution of these galaxies. On the other hand, there is also a clear trend when looking at the \emph{recent} ($\sim0.5-1$ Gyr) level of activity. Namely, dwarfs within a denser region of the group appear to have had their star formation quenched while dwarfs located in the group outskirts show a wide range of possible star formation rates, thus indicating that external processes play a fundamental role, complementary to mass, in shaping the star formation histories of dwarf galaxies.}

\titlerunning{A close look at the Centaurus A group of galaxies IV.}

  \maketitle


\section{Introduction}

The Local Group (LG) is the closest and thus best studied environment in which to analyse the properties of galaxies and their stellar content. In particular, the most numerous population in our own group and more generally in the Universe is that of dwarf galaxies. Given their small sizes and their morphological diversity, they are an ideal laboratory for investigating star formation processes as well as seeking environmental signatures on their evolution. 

While the huge amount of information coming from LG dwarfs has already broadened our understanding of their physical properties and evolutionary histories, in the past two decades we have extensively exploited the capacities of the most advanced telescopes and instruments in order to cross the LG boundaries and study in unprecedent detail more distant galaxies. In particular, dwarf galaxies in nearby groups have been predominantly found to have similar properties to those of our close neighbors \citep[e.g.][]{kara02_m81, kara02, kara03_sc, kara03_can, kara04, kara07, trentham02, kara05, sharina08, bouchard08, weisz08, cote09, dalcanton09, koleva09, mcquinn09, mcquinn10, crnojevic10, crnojevic11a, lianou10, makarova10}. A recent Hubble Space Telescope (HST) survey of dwarf galaxies \citep[ANGST,][]{weisz11}, comprising about 60 objects located within $\sim4$ Mpc, has led to a large compilation of color-magnitude diagrams (CMDs) of their resolved stellar populations. From these, it is possible to extract their star formation histories (SFHs) and to confirm on large scales the already locally known morphology-density relation for groups \citep[e.g.,][]{einasto74, grebel00esa}. The latter sees the predominantly old and gas-depleted early-type dwarfs (dwarf ellipticals, dEs, and dwarf spheroidals, dSphs) being found closest to their giant hosts ($\lesssim300$ kpc), while the star forming and gas-rich late-type dwarfs (or dwarf irregulars, dIrrs) lie in the groups' outskirts. This evidence suggests that some environmental process (e.g., tidal interaction, ram pressure stripping, or a combination of both) might deprive dSphs from their gas content thus stopping star formation, or even transforming them from dIrrs into early-type objects \citep[e.g.][]{grebel03, mayer06, kazantzidis11}. The low-mass transition-type dwarfs present stellar population characteristics similar to those of dSphs (i.e., predominantly old), but also contain some neutral gas, just as dIrrs, with the difference that they do not currently form stars. These dwarfs may be thus examples of galaxies currently experiencing such an evolution. For a more detailed characterization of these different galaxy types, see \citet{grebel01}.

\begin{table*}
 \centering
\caption{Fundamental properties of the CenA dwarf irregular companions.}
\label{infogen}
\begin{tabular}{lcccccccccc}
\hline
\hline
Galaxy&RA&DEC&$D$&$(m-M)_{0}$&$D_{CenA}$&$\theta$&$A_{I}$&$M_{B}$&$M_{HI}$&$\Theta$\\
&(J2000)&(J2000)&(Mpc)&&(kpc)&($^{\circ}$)&&&$(10^6$M$_{\odot})$\\
\hline
\object{KK182, Cen6}&$13\,05\,02.9$&$-40\,04\,58$&$5.78\pm0.42$&$28.81\pm0.15$&$2048\pm555$&$4.82$&$0.20$&$-12.48$&$42/55$\tablefootmark{a,d}&$-0.5$\\
\object{ESO269-58}&$13\,10\,32.9$&$-46\,59\,27$&$3.80\pm0.29$&$27.90\pm0.16$&$316\pm44$&$4.76$&$0.21$&$-14.60$&$24$\tablefootmark{c}&$1.9$\\
\object{KK196, AM1318-444}&$13\,21\,47.1$&$-45\,03\,48$&$3.98\pm0.29$&$28.00\pm0.15$&$255\pm391$&$2.15$&$0.16$&$-11.90$&$-$&$2.2$\\
\object{HIPASS J1348-37}&$13\,48\,33.9$&$-37\,58\,03$&$5.75\pm0.66$&$28.80\pm0.24$&$2053\pm738$&$6.69$&$0.15$&$-11.90$&$8/33$\tablefootmark{a,c}&$-1.2$\\
\object{ESO384-16}&$13\,57\,01.6$&$-35\,20\,02$&$4.53\pm0.31$&$28.28\pm0.14$&$1038\pm352$&$9.81$&$0.14$&$-13.17$&$4/6$\tablefootmark{b,d}&$-0.3$\\
\hline
\end{tabular}
\tablefoot{
The columns are the following: (1): name of the galaxy; (2-3): equatorial coordinates from \citet{kara07} (J2000; units of right ascension are hours, minutes and seconds, and units of declination are degrees, arcminutes and arcseconds); (4-5): distance and distance modulus of the galaxy derived by \citet{kara07} with the tip of the red giant branch method; (6-7): deprojected and angular distance from CenA; (8): Galactic foreground extinction in the $I$-band from \citet{schlegel98}; (9): absolute $B$ magnitude (converted from \citealt{bouchard08} with the distance modulus listed in column (6)); (10): HI mass obtained from different sources (\tablefoottext{a}{\citealt{banks99}, recomputed with updated values of the distance}; \tablefoottext{b}{\citealt{beaulieu06}}; \tablefoottext{c}{\citealt{georgiev08}}; \tablefoottext{d}{\citealt{bouchard08}}); and (11): tidal index (i.e., degree of isolation), taken from \citet{kara07}.
}
\end{table*}

Complementing the ANGST survery, we have started an analysis of the dwarf population residing in the Centaurus A (CenA)/M83 group of galaxies. This dense and dynamically evolved conglomeration is the only group within the Local Volume dominated by a giant elliptical and a giant spiral, thus providing crucial clues on how their small mass companions adapt to a very massive host/a relatively dense environment (with respect to, e.g., the Local Group). The CenA/M83 group has an average distance of $\sim4$ Mpc, extending out to $\sim5.5$ Mpc from us, thus lying slightly further away than the ANGST sample. The first systematic HST study of the resolved stellar populations of its dwarf members was carried out by \citet{kara02}, followed by more recent, even deeper HST observations \citep{kara07}. In \citet{crnojevic11b} we derive recent ($\lesssim 1$ Gyr) SFHs for five dIrr members of the M83 subgroup with the synthetic CMD modeling technique \citep[e.g.,][]{dolphin02, cole07, cignoni10}. In the present paper, we extend our study to five members of the CenA subgroup, with the goal of completing the analysis of the known dwarf galaxies in this group and of seeking possible environmental effects on their evolution.

This paper is organized as follows: in Sect. \ref{data_sec} we describe the data and the photometric reduction process, while in Sect. \ref{cmd_sec} we show the resulting CMDs. Sect. \ref{sfh_sec} reports the derived SFHs for each galaxy, and Sect. \ref{maps_sec} shows the spatial distribution of their different stellar subpopulations. We discuss our findings in Sect. \ref{discuss}, and finally draw the conclusions in Sect. \ref{conclus}.


\section{Data and photometry} \label{data_sec}

We use archival data from the observing programmes GO-9771 and GO-10235, taken with the the Wide Field Channel (WFC) of the Advanced Camera for Surveys (ACS) aboard the Hubble Space Telescope (HST). Each of our five targets was observed in the F606W (corresponding to the broad $V$-band in the Johnson-Cousins system, 1200 seconds exposure) and in the F814W filters (broad $I$-band, 900 seconds exposure).

The photometry was performed using the ACS module of the DOLPHOT package \citep{dolphin02}, adopting the parameters suggested in the User's Guide\footnote{http://purcell.as.arizona.edu/dolphot/.}. The output catalogs contain magnitudes in both the ACS photometric system bands and magnitudes transformed to the Johnson-Cousins system bands. The quality selection criteria applied to the photometry are the following: signal-to-noise $\geq5$, $\chi^2\leq2.5$, sharpness between $-0.3$ and 0.3, and crowding parameter $\leq0.5$ in both filters. We note that for some of the target galaxies (ESO384-16 and ESO269-58) the central regions are highly crowded and the above-mentioned quality cuts would reject good stars in this region. To avoid this, we change the Force1 parameter in the photometry from the suggested value in order not to discard a priori sources that are too faint, too sharp or too elongated. We afterwards apply stricter quality cuts ($-0.2\leq$ sharpness $\leq0.2$) in order to separate bona fide stars from spurious objects. For more details about the quality parameters and cuts, we refer the reader to the explanation in \citet{crnojevic11b}.

We perform artificial star tests in order to accurately estimate the photometric errors and the completeness curves for our galaxies, and in order to constrain the star formation parameters derived in the SFH recovery process (see Sect. \ref{sfh_sec}). We add about $5-10$ times more artificial stars to each image than the number of galaxy stars (after quality cuts), evenly covering the ACS field of view and the color-magnitude space spanned by the galaxy stars. We moreover add artificial stars up to 1 mag below the faintest galaxy stars in order to account for objects that are upscattered in the CMD due to an addition of noise. As an example we give the derived photometric errors and completeness for the least and most crowded galaxies in this sample, i.e., HIPASS J1348-37 (with a peak density of $\sim$ two stars per arcsec$^{2}$, corresponding to $\sim184$ stars per $0.1$ kpc$^{2}$) and ESO269-58 (peak density of $\sim 10$ stars per arcsec$^{2}$, or $\sim2520$ stars per $0.1$ kpc$^{2}$) respectively. For HIPASS J1348-37, the limiting magnitude at a $50\%$ completeness level is $\sim\!27.0$ mag ($\sim\!26.0$ mag) for the $V$-band ($I$-band). At the mentioned $I$-band magnitude, the representative $1 \sigma$ photometric error amounts to $\sim\!0.21$ mag in magnitude and $\sim\!0.27$ mag in color. For ESO269-58, the $50\%$ completeness level is reached already at $\sim\!26.2$ mag ($\sim\!25.3$ mag) for the $V$-band ($I$-band), while the corresponding $1 \sigma$ photometric errors for this $I$-band magnitude are $\sim\!0.20$ mag in magnitude and $\sim\!0.28$ mag in color. Photometric errors are shown for each galaxy in the CMDs of the next Section.

The general properties of the studied CenA dwarf irregular companions are reported in Table \ref{infogen}. Note that all the galaxies have a morphological de Vaucouleurs type of 10.

\begin{figure*}
 \centering
  \includegraphics[width=17.5cm]{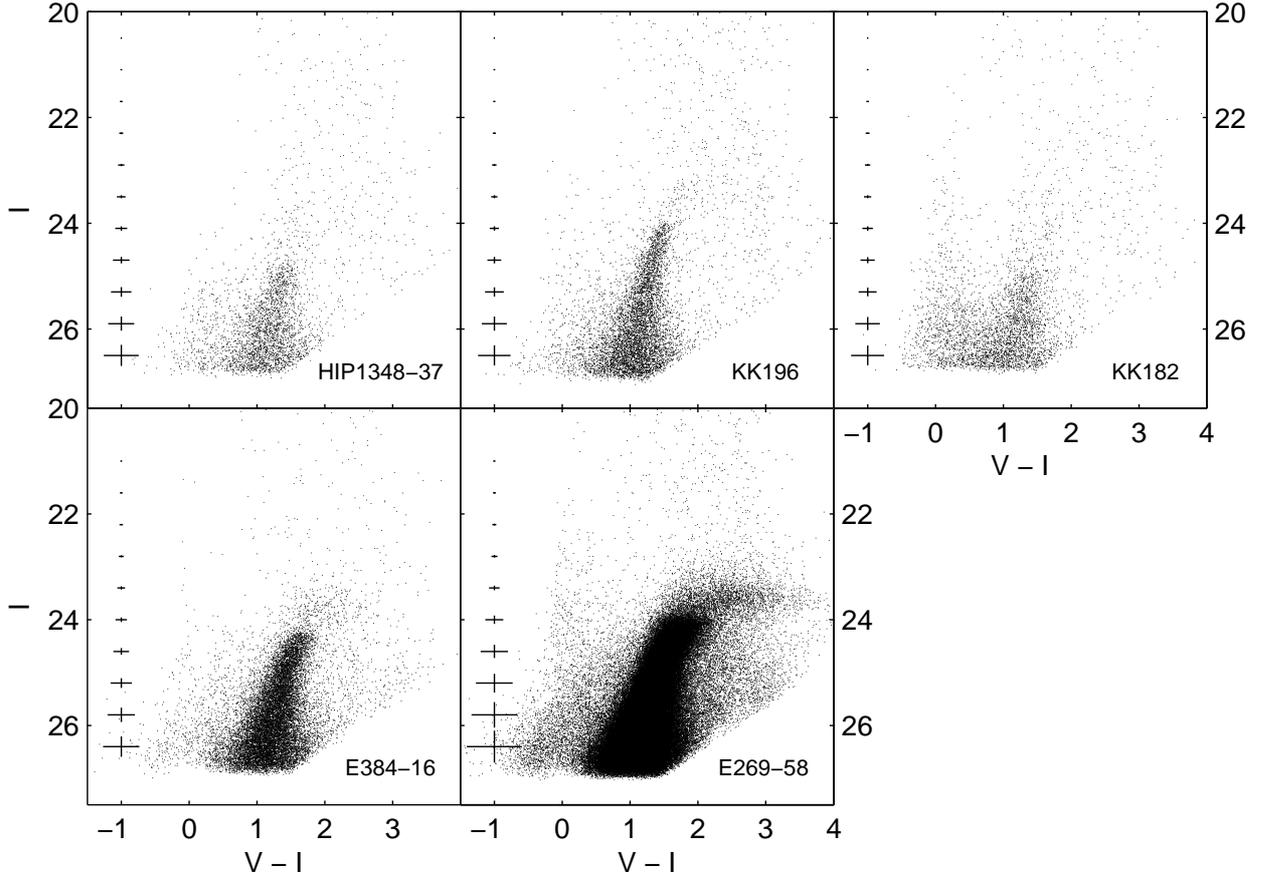}
\caption{\footnotesize{Color-magnitude diagrams of the five late-type dwarf companions of CenA, ordered by (increasing) luminosity. The main features visible in all of the diagrams are the blue plume (main sequence and blue helium-burning stars, very sparsely populated for HIPASS J1348-37, KK196, and ESO384-16), the red supergiant phase (or red helium-burning stars, significant only for KK182), the upper red giant branch and the luminous asymptotic giant branch (see text for details). On the left side of each diagram, representative $1\sigma$ photometric errorbars (as derived from artificial star tests) are shown.}}
 \label{fore}
\end{figure*}


\section{Color-magnitude diagrams} \label{cmd_sec}

In Fig. \ref{fore} we show the CMDs for our target galaxies, ordered by increasing absolute magnitude. The photometric errors are indicated as representative errorbars along the CMDs. The $1 \sigma$ error has a value of $\sim0.1$ mag in magnitude and $\sim0.15$ mag in color at different $I$-band magnitudes for different galaxies, namely: $25.40$ for HIPASS J1348-37, $25.40$ for KK196, $25.55$ for KK182, $25.30$ for ESO384-16, and $24.30$ for ESO269-58.

\begin{figure}
 \centering
  \includegraphics[width=9cm]{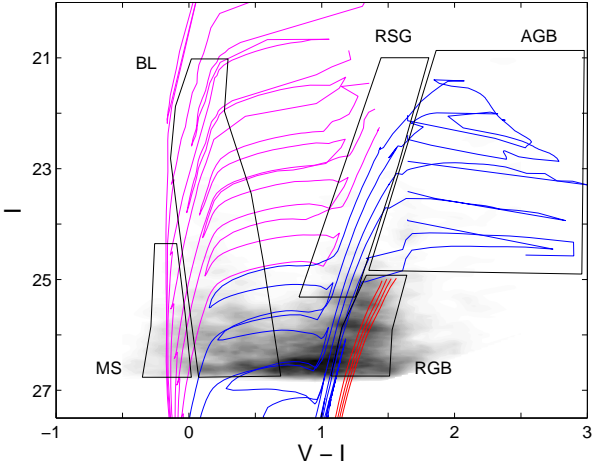}
\caption{\footnotesize{Hess density diagram of the dwarf irregular galaxy KK182. Overlaid are Padova stellar isochrones, with a fixed metallicity of Z=0.0008 and varying ages. Ages of 4, 8, 15, 20, 35, 55 and 85 Myr (proceeding from the blue to the red part of the CMD) are drawn in magenta, ages of 130, 200, 350, 550, 800 Myr and 1.3 Gyr are in blue, and ages of 4, 7, 10 and 14 Gyr are in red (see text for details). Also plotted are selection boxes that separate the different evolutionary stages, as indicated.}}
 \label{isos}
\end{figure}

In Fig. \ref{isos} we show a Hess diagram for the galaxy KK182 and to it we overlay Padova isochrones \citep{marigo08} in order to highlight the main evolutionary features present in the CMD. We choose isochrones with a fixed metallicity of Z=0.0008 (corresponding to [Fe/H]$\sim-1.4$ when assuming Z$_\odot$=0.019) and ages varying from 4 Myr to 14 Gyr. The chosen metallicity is the average value obtained from our best-fit SFH for this galaxy (see next section). The visible features are: the upper part of the red giant branch (RGB) in the range $V-I\sim1$ to 1.6, indicative of intermediate-age and old populations ($\geq 1-2$ Gyr); the luminous asymptotic giant branch (AGB) at $1.3\leq V-I\leq2.5$, above the tip of the RGB (TRGB), i.e. low-mass intermediate-age stars burning helium in their core; a sequence of massive helium-burning stars with ages from $\sim20$ to 500 Myr, moving from the blue ($V-I\sim0$ to 0.5, ``blue-loop'' or BL stars) to the red ($V-I\sim1$ to 1.3, red supergiant phase, RSG) part of the CMD during their evolution; and massive upper main sequence (MS) stars as young as $\sim10$ Myr, found at $-0.5\leq V-I \leq0$ and $I\geq23$. We also draw boxes to indicate the selection made when separating the different evolutionary stages (see Sect. \ref{sfh_sec}), which is based on the mentioned evolutionary sequences present in each CMD and on the Padova isochrones. We note that the models do not always perfectly represent the observed evolutionary sequences, which is particularly true for the RSG phase \citep[e.g.,][and references therein]{mcquinn11}. As already mentioned in \citet{crnojevic11b}, the observed RSG sequence is in fact redder than the models, and we thus draw the selection box accordingly to the observed sequence.

As already extensively discussed in \citet{crnojevic11b}, at an average Galactic latitude of $b\sim\!20^{\circ}$ the foreground stellar contamination from Milky Way stars is not negligible. This effect can be estimated by simulating the Galactic foreground stars with theoretical models as, for example, the TRILEGAL models \citep{girardi05}, and then overplotting their expected distribution on to the CMDs of the target dwarf galaxies (see, e.g., Fig. 2 of \citealt{crnojevic11b}). These stars fall on a region of our observed CMDs that encompasses colors from $V-I\sim0.5$ to 3.5, while the most affected part will be the poorly populated RSG region. In particular, the relative contamination for the galaxies studied here varies from $10\%$ up to $20\%$ of the RSG stars. We thus perform a statistical foreground decontamination on our CMDs prior to running our SFH recovery code, following the method described in \citet{crnojevic11b}. Briefly, we produce one random realization of the possible foreground contamination with the TRILEGAL model and we take into account the incompleteness effects and the photometric errors stemming from the observations on the simulated sample of stars. For each of the Galactic stars we then subtract one star in the observed CMD of the dwarf galaxy, corresponding to its position in the color-magnitude space.

Once the foreground contamination has been taken into account, our purpose is to use the information present in the CMDs in order to reconstruct the details of the past star formation activity in our galaxies. Considering the number of stars in each different evolutionary stage to the extent of information given by the CMDs, i.e. RGB, AGB, BL, RSG and MS, with the help of theoretical stellar population models we can constrain the \emph{SFR} at different epochs by using a dedicated code (developed by A. A. Cole), as explained in the next section.

\section{Star formation histories} \label{sfh_sec}

\begin{table*}
 \centering
 \caption{\footnotesize{Star formation parameters and metallicity derived for the studied galaxies.}} 
 \label{ressfh}
\begin{tabular}{lcccccccccc}
\hline
\hline
Galaxy&$<SFR>$&$b_{100}$&$b_{500}$&$b_{1G}$&$b_{14G}$&$f_{1G}$&$f_{4G}$&$f_{14G}$&$<$[Fe/H]$>$&$M_{star}$\tablefootmark{a}\\
&($10^{-2}$M$_{\odot}$yr$^{-1})$&&&&&&&&(dex)&($10^{7}$M$_{\odot}$)\\
\hline
\object{HIPASS J1348-37}&$0.14\pm0.05$&$1.56$&$1.78$&$1.73$&$0.88$&$0.12$&$0.28$&$0.60$&$-1.50\pm0.07$&$1.9^{+1.1}_{-1.0}$\\
\object{KK196, AM1318-444}&$0.20\pm0.11$&$-$&$0.99$&$0.66$&$1.04$&$0.07$&$0.19$&$0.74$&$-1.43\pm0.25$&$2.7^{+0.6}_{-0.9}$\\
\object{KK182, Cen6}&$0.10\pm0.14$&$5.72$&$6.78$&$3.37$&$0.85$&$0.24$&$0.35$&$0.41$&$-1.46\pm0.21$&$1.3^{+1.0}_{-0.7}$\\
\object{ESO384-16}&$0.60\pm0.23$&$0.14$&$0.54$&$0.37$&$1.09$&$0.03$&$0.27$&$0.70$&$-0.97\pm0.15$&$8.0^{+2.0}_{-2.5}$\\
\object{ESO269-58}&$6.79\pm3.93$&$0.12$&$0.31$&$0.21$&$1.11$&$0.01$&$0.22$&$0.76$&$-0.98\pm0.20$&$88.0^{+11.0}_{-23.0}$\\
\hline
\end{tabular}
\tablefoot{
The columns are the following: column (1): name of the galaxy (ordered by increasing absolute magnitude); (2): average \emph{SFR} over the whole lifetime; (3): $b_{100}$; (4): $b_{500}$; (5): $b_{1G}$; (6): $b_{14G}$; (7): $f_{1G}$; (8): $f_{4G}$; (9): $f_{14G}$; (10): average metallicity over the whole lifetime; and (11): stellar mass as integrated from our best-fit SFH (\tablefoottext{a}{note that these are possible lower limits because of incomplete spatial coverage; moreover, these estimates depend on the modeling details, and do not take into account stars that already ended their lives}). The $b$ parameter is the ratio of star formation rate over the indicated time period (0.1, 0.5, 1 and from 1 to 14 Gyr respectively) to the average star formation over the whole lifetime; the $f$ parameter is the fraction of stars born in a certain time interval ($0-1$, $1-4$ and $4-14$ Gyr). In a hypothetical galaxy with constant star formation rate, $f_{1G}=0.07$, $f_{4G}=0.36$ and $f_{14G}=0.57$.
}
\end{table*}

\begin{figure*}
 \centering
  \includegraphics[scale=0.57]{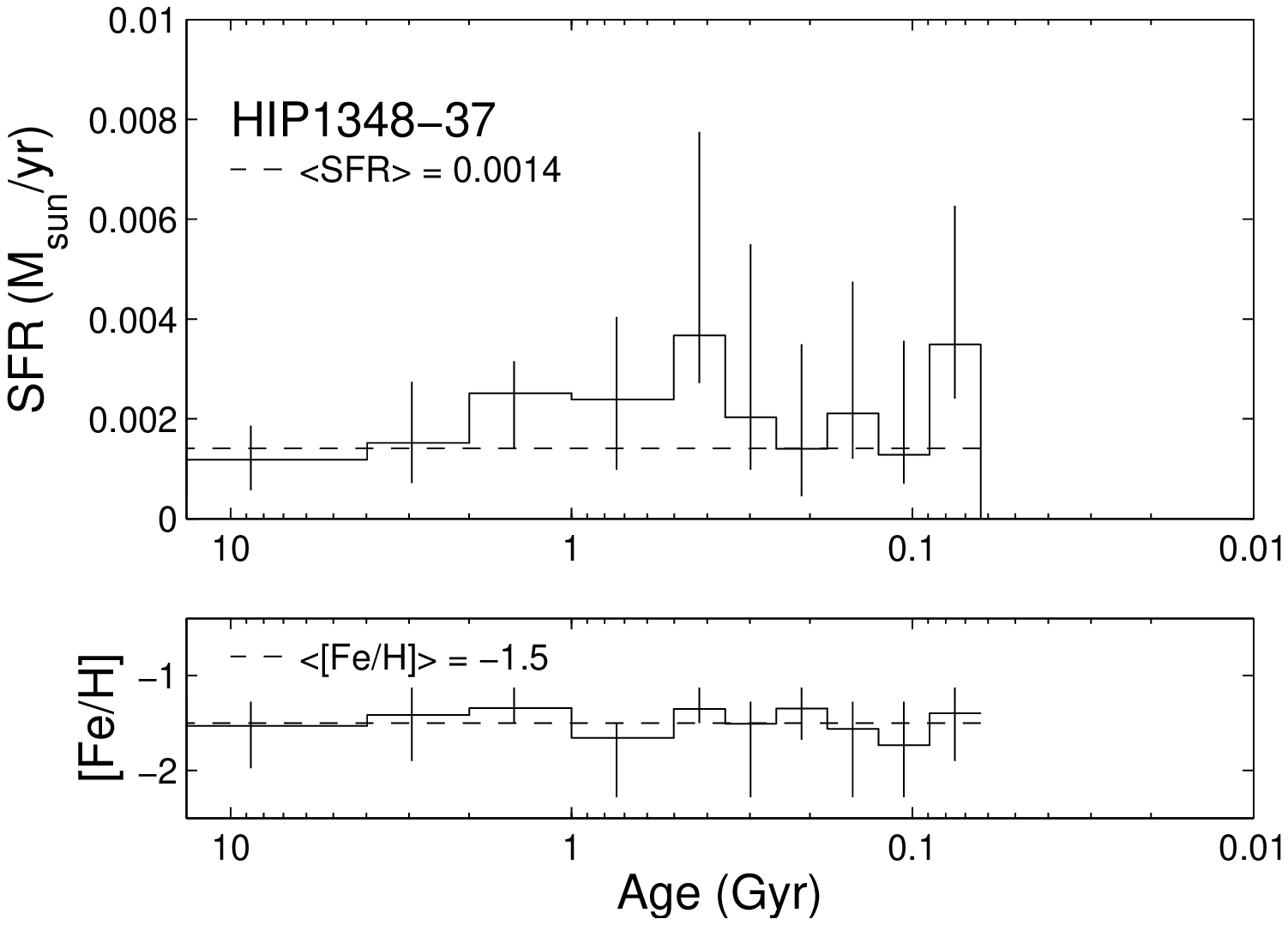}
  \includegraphics[scale=0.57]{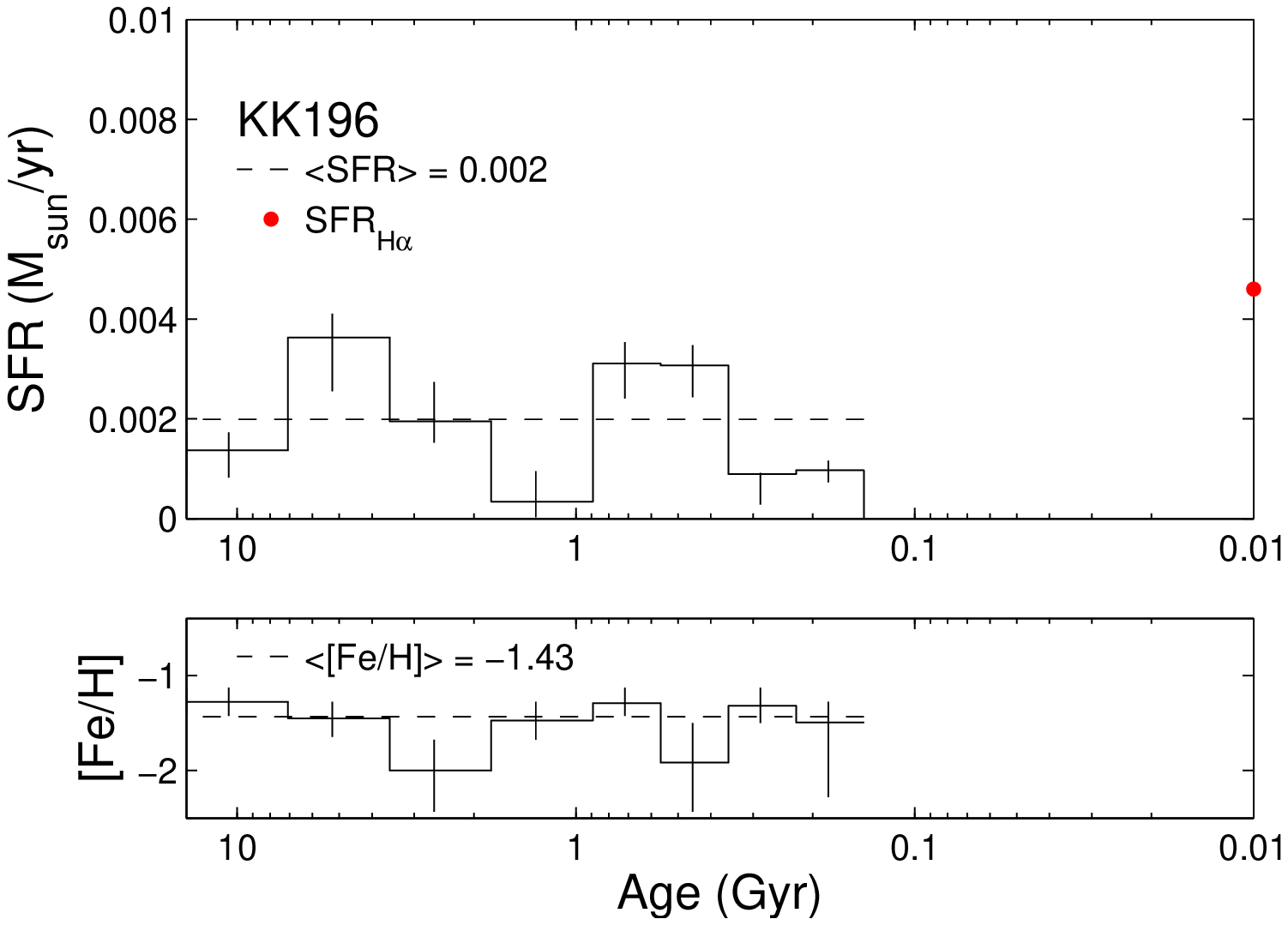}
  \includegraphics[scale=0.57]{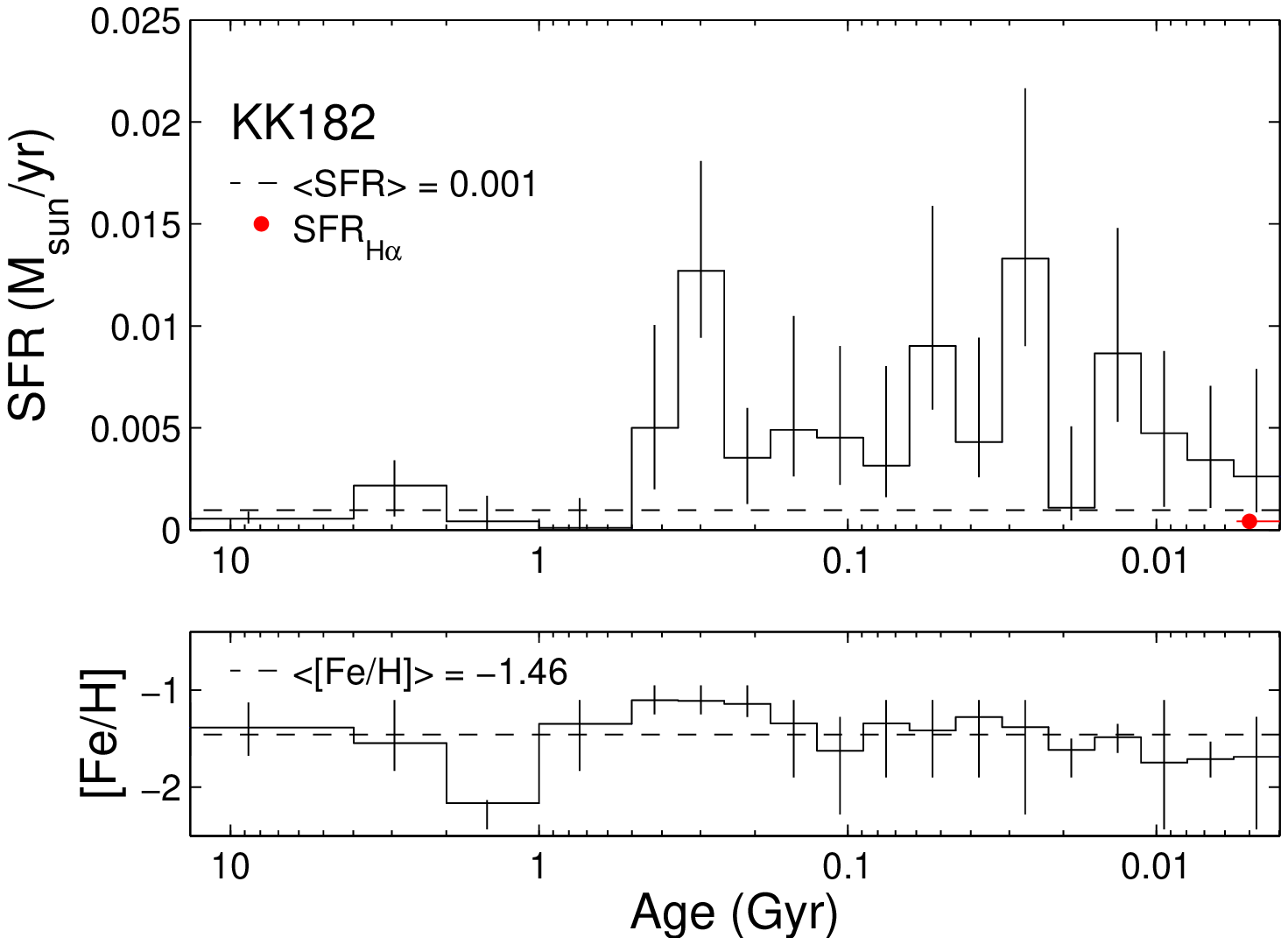}
  \includegraphics[scale=0.57]{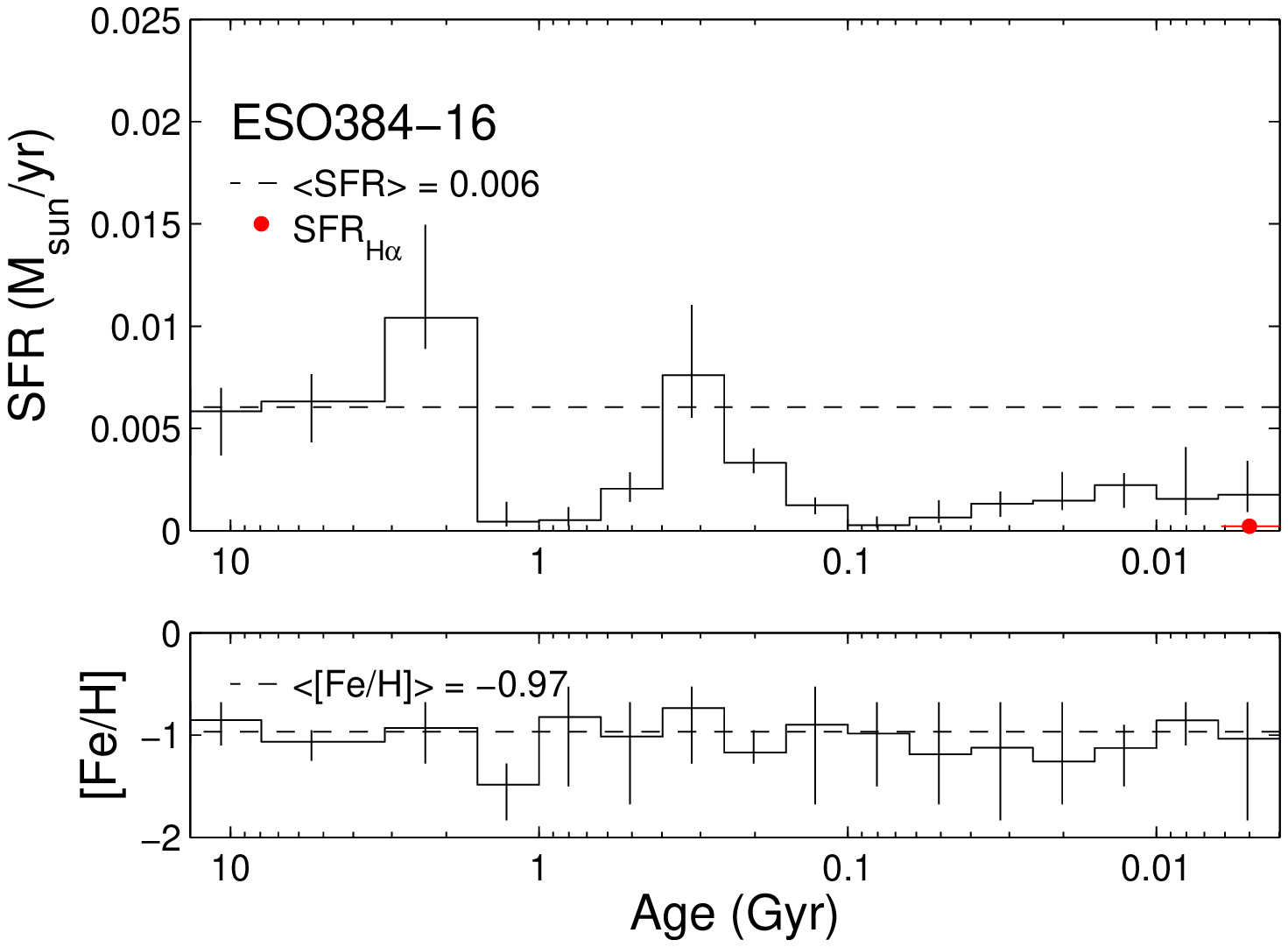}
  \includegraphics[scale=0.57]{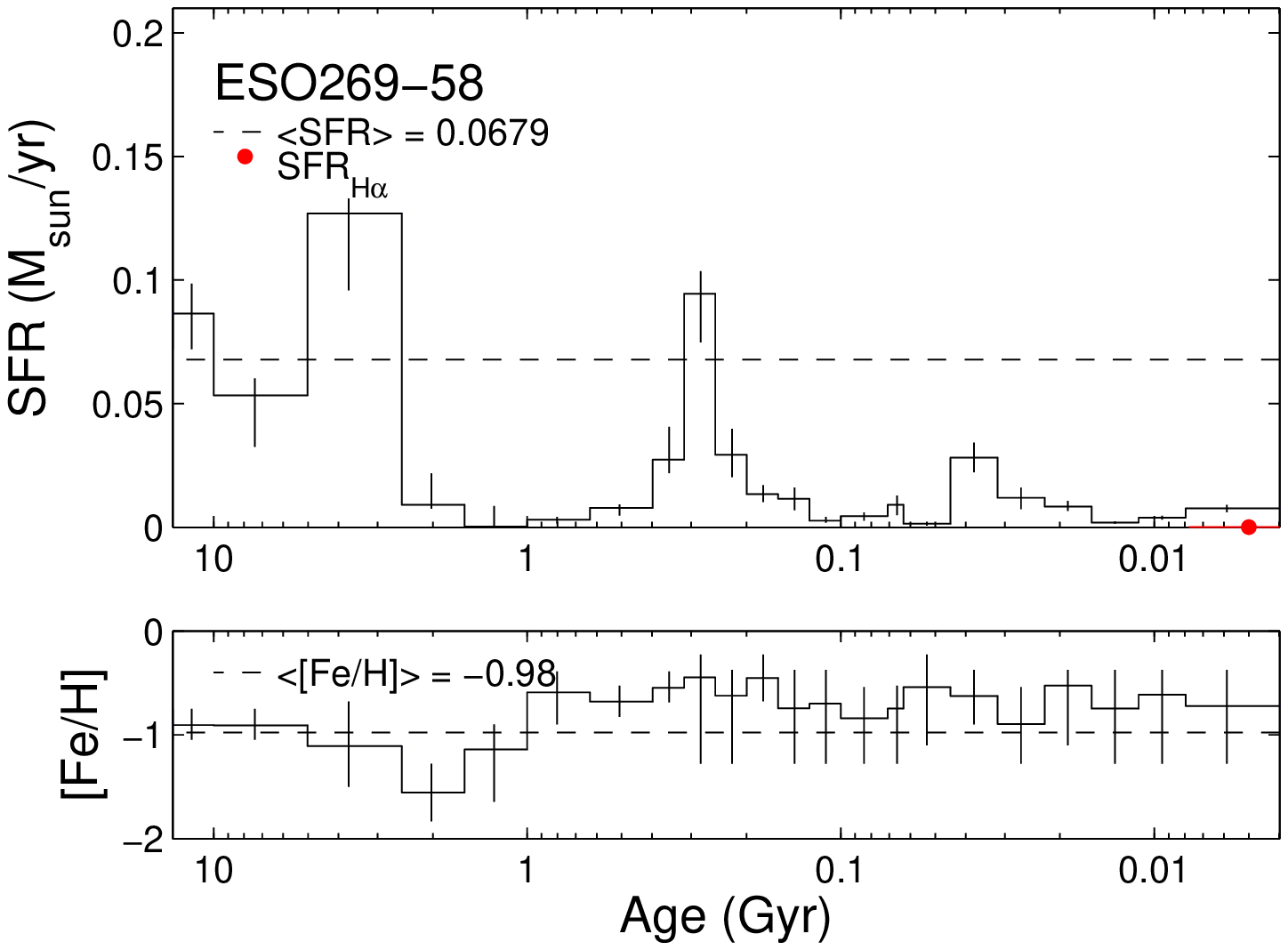}
 \caption{\footnotesize{\emph{Upper panels}. Star formation histories derived for the five studied galaxies (HIPASS J1348-37, KK196, KK182, ESO384-016, and ESO269-58, ordered by increasing absolute magnitude). For each galaxy, the star formation rate as a function of time is plotted, with the oldest age being on the left side and the most recent age bin on the right edge of the (logarithmic) horizontal axis. Note that the size of the age bins is variable due to the different amount of information obtainable from each CMD for each stellar evolutionary stage. For HIPASS J1348-37 and KK196, there is not enough information in the CMD to derive the \emph{SFR} for ages in the past $\sim100$ Myr). For KK182, ESO384-016, and ESO269-58, the CMDs allow us to derive the \emph{SFR} for ages younger than 10 Myr. Also note the different vertical axis scales. The black dashed line indicates the mean star formation rate over the galaxy's lifetime. The red dots (for all the galaxies but for HIPASS J1348-37) report the current star formation rate derived in the literature from their H$\alpha$ emission (often confined to a few individual HII regions). \emph{Lower panels}. Metallicity as a function of time, with the same x-axes as above. The black dashed line represents the mean metallicity over the galaxy's lifetime. Note that the metallicity evolution is poorly constrained.}}
 \label{sfhsnew}
\end{figure*}

The technique we used to derive SFHs for our targets was already extensively described in \citet{crnojevic11b}. Here, we briefly summarize the main points of the SFH derivation process.

The method we use follows the general approach of the well-established so-called synthetic CMD modeling \citep[e.g.][and references therin]{gallart05, tolstoy09}. Starting from Padova stellar evolutionary models \citep{marigo08}, the adopted code builds synthetic CMDs with a wide range of ages and metallicities. We moreover make assumptions about the intial mass function (IMF) and the binary fraction. In particular, the chosen IMF is log-normal between 0.08 and 1 M$_\odot$ \citep{chabrier03} and it follows a Salpeter power-law above 1 M$_\odot$. For the binary fraction, prescriptions from \citet{duque91} and \citet{mazeh92} are used to assign a companion star drawn from the same IMF to $65\%$ of the primary stars. Distance moduli and reddening values are taken from \citet{kara07} (derived via the TRGB method) and from the Schlegel extinction maps \citep{schlegel98}, respectively. 

We do not make any assumption about a possible age-metallicity relation. This would, in fact, not change the amount of star formation, but it would slightly affect the age and duration of each star formation episode. Due to the limited amount of information available from the CMDs, any attempt to impose such a relation would be ``artificial''. We initially choose 5 different isochrone metallicities to construct the synthetic CMDs ($Z=0.0001, 0.00024, 0.0004, 0.0006$ and $0.001$). We have tested the possibility of using higher metallicity values (see \citealt{crnojevic11b}), but while this increases both the computational time and the uncertainties in the final metallicity results, it does not appear to significantly affect the resulting SFH. For only one galaxy (KK196), a nebular oxygen abundance value is available from the literature (see KK196 subsection), and we use that to constrain its present-day iron abundance.

As shown in \citet{crnojevic11b}, assuming an internal reddening a priori is not necessary, as the code lets the initial input value vary slightly until the best solution is found. We also chose not to take into account differential reddening internal to the galaxies. This is almost certainly present, as evidenced by the presence of (non-uniformly distributed) young stars. However, \citet{cignoni10}, in their review paper about star formation modeling techniques, show that the effects of not taking differential reddening into account for the SFH reconstruction are minor and not systematic, as compared for example to the adopted foreground reddening value. Moreover, since we are dealing with metal-poor dwarf galaxies, the amount of dust is generally low. Finally, in \citet{crnojevic11b} we tested the effects of an increased internal reddening on one of our targets, concluding that it does not significantly affect the results of our SFH recovery.

We make use of the artificial star tests to take into account photometric errors and incompleteness effects in our observations. The synthetic CMDs are then binned (depending on the characteristics of the observed CMD) and the number of stars in each grid cell is computed and compared to the ones in the observed CMD. We can now look for the synthetic SFH producing a CMD that has the maximum likelihood to match the observed one. This is done by exploring the parameter space via a simulated annealing algorithm \citep[described in ][]{skillman03, cole07}. This procedure gives for each galaxy the \emph{SFR} at each chosen time interval, and a corresponding best-fit metallicity value. The solution will be mostly driven by the CMD grid cells that are most populous, thus giving a robust result as long as the incompleteness effects are carefully taken into account. The vertical errorbars are then obtained by perturbing the best-fit \emph{SFR} value for an individual age bin, and re-running the global SFH recovery process. The individual age bin \emph{SFR} value is varied until the global SFH solution stays within 1$\sigma$ of the global best-fit value, which sets the errorbar on that specific age bin. Subsequently, the process is repeated starting from the other age bins. The best-fit SFHs and age-metallicity relations are shown in Fig. \ref{sfhsnew}.

Note that in the resulting SFHs the age bins are approximately logarithmic, as is typical for SFH derivations by synthetic CMDs, due to a progressive lack of information in the CMDs for increasing ages. The size of the bins is ultimately limited by the age resolution of the data (as discussed in, e.g., \citealt{hidalgo11}), but at young ages depends on the star formation rate and the mass-luminosity relation (i.e., the variation in age across the color-magnitude bins and number of stars in each bin). The choice of when to change the size of the bins in (logarithmic) age is based on the appearance of different age-sensitive features in the CMD (e.g., MS turnoffs and BL stars). As shown in \citet{cole07} and discussed in \citet{aparicio09}, SFH solutions obtained in this way are generally robust to minor variations in the exact binning scheme adopted. Our CMDs are not deep enough to include features like the old MS turnoff, the red clump or the horizontal branch, due to the distance of our targets, and it is thus difficult to accurately constrain the star formation episodes for ages $\gtrsim1$ Gyr from the evolved stellar populations alone. In particular, the number of AGB and RGB stars will constrain the total \emph{SFR} at these ages, but it is not possible to resolve single bursts of star formation older than 1 Gyr. For the oldest populations the choice of wider age bins is partly motivated by the desire to suppress the model-dependent covariance arising from the age-metallicity degeneracy. This covariance is accounted for by the process of generating the errorbars. Given the age-metallicity degeneracy present at these ages, also the metallicity evolution with time cannot be accurately constrained, although the derived lifetime mean values are robust and in no case there seems to be a strong increase of metallicity with time (consistent with results of dwarf irregulars in the Local Group, e.g., \citealt{grebel04a, glatt08b}). Finally, the errorbars on the metallicities shown in Fig. \ref{sfhsnew} represent the range from the $10^{th}$ to the $90^{th}$ percentile from the best-fit SFH.

In addition to the mean values of \emph{SFR} and metallicity for each galaxy, in Tab. \ref{ressfh} we report the parameters that characterize the derived SFHs, in order to be able to compare our results to the M83 dwarfs subsample and to other literature studies. In particular, the $b$ parameter gives the ratio of star formation rate over a certain time period to the average star formation over the whole lifetime ($b_{100}$ over the last 100 Myr, $b_{500}$ over the last 500 Myr, $b_{1G}$ over the last 1 Gyr, and $b_{14G}$ for ages older then 1 Gyr ago). The so-called birthrate parameter is defined to be the fraction of stars formed from $0-1$ Gry ($f_{1G}$), from $1-4$ Gyr ($f_{4G}$) and from $4-14$ Gyr ($f_{14G}$). In the same table, we also report the integrated stellar masses from our best-fit SFHs (see also following subsections). Note that, firstly, these masses only refer to the area covered by the ACS field-of-view (FoV), while in some cases the faint outskirts of our targets could be extending beyond that. We choose not to correct these values with a simplistic area normalization factor, as we do not know the real extent of our targets, and as this would not take into account possible spatial variations of the SFH. Secondly, the derived masses suffer from uncertainties intrinsic to the SFH modeling process, e.g., the choice of the IMF. Finally, the computed values could systematically overestimate the true one by up to $\sim\!25\%$, because we do not take into account stars that already ended their lives.

In the following subsections we report and discuss the detailed results of the SFH recovery code for each individual galaxy of this sample.

\subsection{HIPASS J1348-37} \label{}

Within the CenA subgroup of late-type dwarfs, HIPASS J1348-37 is the faintest one (together with KK196, see Table~\ref{infogen}), and is the galaxy with the lowest density. Due to its large deprojected distance from CenA ($\sim\!2.1\pm0.7$ Mpc, computed using the radial distances found in \citealt{kara07}), we also try to compute its deprojected distance from M83, and find that HIPASS J1348-37 is actually closer to the latter than to CenA ($\sim\!1.0\pm0.5$ Mpc). The subgroup membership of our late-type targets was initially assigned following the classification of \citet{kara05}. However, with the new and more precise distance measurements of \citet{kara07}, HIPASS J1348-37 (together with other two galaxies of this subsample, namely KK182 and ESO384-16) is found to have a negative tidal index ($-1.2$, \citealt{kara07}) instead of the positive one that was reported in older papers, and thus turns out to be a rather isolated galaxy. In our subsequent analysis, we will consider this information, as explained more in detail in the discussion section. 

HIPASS J1348-37 contains a few $\sim\!10^{7}$M$_\odot$ of neutral gas \citep{banks99, georgiev08}, but has not been studied in H$\alpha$ yet. Its CMD, presented in Fig.~\ref{fore} and containing $\sim\!3300$ stars, shows an absence of MS stars and presence of a number of BL, RSG and luminous AGB stars. As ESO444-78 is in the M83 sample, from the characteristics of its CMD HIPASS J1348-37 could be classified as transition-type dwarf, i.e. it contains $>10^6$M$_\odot$ of neutral gas but shows an absence of star formation activity in the past $\sim60$ Myr \citep[e.g.][]{grebel03}.

We derive the SFH for HIPASS J1348-37, although the information stemming from the CMD is mainly based on the old RGB stars and on a few intermediate-age stars, and plot it in Fig.~\ref{sfhsnew}. From the lack of bright MS and BL stars, we find no significant star formation for ages younger than $\sim\!60$ Myr, and the average \emph{SFR} from ages older than that is $\sim\!0.0014\pm0.0005$M$_\odot$yr$^{-1}$. As in \citet{crnojevic11b}, we adopt a standard $\Lambda$CDM cosmology with $t_0=13.7$ Gyr, $H_0=71$ km s$^{-1}$ Mpc$^{-1}$, $\Omega_{\Lambda}=0.73$ and $\Omega_{m}=0.27$. Assuming the formation epoch of the galaxy to have been around $\sim\!13.5$ Gyr ago, we thus estimate that HIPASS J1348-37 had already formed $\sim 50\% ^{+10\%}_{-40\%}$ of its stellar content prior to 5 Gyr ago ($z\sim\!0.5$). Note that the errorbars reflect the age uncertainty in the SFH at these old ages. In particular, this galaxy has been slightly more active in its last $\sim\!1$ Gyr of life, as also indicated by the parameters listed in Table~\ref{ressfh}, with moderate enhancements of the star formation $\sim\!80$ and $\sim\!400$ Myr ago. The star formation episode at $\sim\!400$ Myr ago is very uncertain, because this is the oldest age where the blue loop stars are included in the CMD and it may indicate increased noise at the lower end of the CMD instead of a real increase in star formation. However, the enhancement in the $\sim\!250-1000$ Myr range is real because of the large number of stars on the blue side of the RGB (they are too blue to be ancient, metal-poor stars).

The average metallicity of HIPASS J1348-37 derived from the SFH recovery is [Fe/H]$=-1.50\pm0.07$. As a test of our derived SFHs, we compute the total stellar mass for the galaxy in two different ways. We find a value of $1.9^{+1.1}_{-1.0} \times10^{7}$M$_\odot$ from our SFH. On the other hand, we consider the $B$-band luminosity and assume a stellar mass-to-light ratio of 1 or 2 (typical values for gas-rich dwarfs, e.g., \citealt{banks99}). We can then estimate the stellar mass to be $0.9\times10^{7}$M$_\odot$ to $1.7\times10^{7}$M$_\odot$, which is consistent with the previously derived value.

\subsection{KK196, AM1318-444} \label{}

As faint as HIPASS J1348-37, but with a higher stellar density, KK196 is the next object considered in our sample, with $\sim\!7300$ stellar sources detected. Although its CMD shows features similar to HIPASS J1348-37 (Fig.~\ref{fore}, few BL and RSG stars, with a small stellar concentration in the luminous AGB region), KK196 is located much closer to CenA, at a deprojected distance of $\sim\!260\pm390$ kpc and in a denser environment ($\Theta=2.2$, \citealt{kara07}). Moreover, its projected position puts it in the middle of CenA's southern radio lobe \citep[see][]{israel98}. KK196 has properties similar to ESO444-78 and IC4247 in the M83 subgroup, however, it is not detected in HI \citep[e.g.,][]{kara07}. Curiously, this dwarf shows almost no sign of a MS in its CMD, but it has an H$\alpha$ detection from \citet{lee07}. Finally, LEDA\footnote{http://leda.univ-lyon1.fr/.} reports an internal extinction value in the $B$-band of 0.50 mag due to the inclination of the galaxy.

We show the result of our SFH recovery in Fig.~\ref{sfhsnew}. As for HIPASS J1348-37, the CMD does not show the presence of stars with ages $\lesssim\!100$ Myr. The average \emph{SFR} is $\sim\!0.002\pm0.0011$M$_\odot$yr$^{-1}$, with an enhancement at $\sim\!400-900$ Myr ago, preceded by $\sim\!1$ Gyr of low activity. In the period $\sim\!3.5-7$ Gyr ago, KK196 also experienced a \emph{SFR} slightly higher than the average one, although at such old ages we do not have the resolution to see smaller fluctuations. Overall, this galaxy has not been very active in the past $\sim300-400$ Myr (see Table~\ref{ressfh}), and the H$\alpha$ detection mentioned above can be the result of a short, localized, burst of star formation, which does not produce enough stars to be significantly detected in the CMD. Indeed, if we look at the HST image of KK196 and check where our detected stars are, the (small) central region where H$\alpha$ was detected by \citet{lee07} is not resolved. From the H$\alpha$ data we are able to compute the current \emph{SFR} for KK196 (for details see the Section about IC4247), which has a value of $\sim\!0.0046\pm0.0004$M$_\odot$yr$^{-1}$, and which we also plot in Fig.~\ref{sfhsnew}. Since this data point refers only to one single star-forming region in this galaxy, it is not representative of the most recent ($\sim10$ Myr) star formation activity of the galaxy as a whole, which lies well below this value. KK196 has formed $\sim 60\% ^{+20\%}_{-30\%}$ of its stars more than 5 Gyr ago.

From the SFH recovery, we derive an average metallicity value of [Fe/H]$=-1.43\pm0.25$, which is in excellent agreement with the value we compute starting from the oxygen abundance listed in \citet{lee07}, [Fe/H]$=-1.43\pm0.02$. If we compute the total stellar mass for KK196 from our SFH, we get a value of $2.7^{+0.6}_{-0.9} \times10^{7}$M$_\odot$. When we assume a stellar mass-to-light ratio of 1 or 2 and compute the total stellar mass starting from the $B$-band luminosity, the values are $0.9\times10^{7}$M$_\odot$ and $1.7\times10^{7}$M$_\odot$, thus slightly higher than, but in good agreement with, our first estimate.

\subsection{KK182, Cen6} \label{}

The CMD for KK182 (Fig.~\ref{fore}) has $\sim\!4100$ stars, and shows a fairly well populated MS/BL region, extending to much younger ages than in HIPASS J1348-37 and KK196, although there are comparatively few RGB and luminous AGB stars. This galaxy has a high deprojected distance from CenA ($\sim\!2.0\pm0.6$ Mpc) but it is actually closer to M83 ($\sim\!1.3\pm0.3$ Mpc), with a negative tidal index that makes it a quite isolated object \citep{kara07}. KK182 contains a moderate amount of neutral gas ($\sim\!5\times10^{7}$M$_\odot$) and has also been detected in H$\alpha$ \citep{cote09}. The internal extinction (due to inclination) given by LEDA is $A_B\sim\!0.78$. Finally, the optical image of KK182 reveals a peculiar triangular shape for this object.

When deriving the SFH for this object, we notice that the intermediate-age and old populations provide very little information (Fig.~\ref{sfhsnew}). The average \emph{SFR} for this dwarf is the lowest of this sample, and has a value of $\sim\!0.001\pm0.0014$M$_\odot$yr$^{-1}$. KK182 has experienced most of its star formation in the past $\sim\!0.5$ Gyr, with many short-lived episodes of star formation ($\sim\!10-100$ Myr in duration) for ages $<0.5$ Gyr, the strongest of which had strengths of several times the average \emph{SFR}. This is reflected in the high values of $b_{100}$, $b_{500}$, and $b_{1G}$ reported in Table~\ref{ressfh}. KK182 thus formed only $\sim 35\% ^{+15\%}_{-25\%}$ of its stars more than 5 Gyr ago. This result is quite similar to the SFH for ESO443-09 in the M83 subgroup \citep{crnojevic11b}, the latter being also a very isolated object. The star formation value we derive for the youngest time bin (resolving ages of $\sim\!4$ Myr) is not in disagreement with the value that \citet{cote09} find starting from the H$\alpha$ flux of this galaxy ($\sim\!0.0004$M$_\odot$yr$^{-1}$, see also Fig.~\ref{sfhsnew}), considering its off-center star forming region. Their study does not give errorbars, but it is likely that the two values are similar once the observational errors are considered.

The metallicity content of this galaxy is low and in line with its luminosity ([Fe/H]$=-1.46\pm0.21$). The stellar mass as obtained from the SFH is $1.3^{+1.0}_{-0.7} \times10^{7}$M$_\odot$. If we compute again the stellar mass from the luminosity of KK182 and a given stellar mass-to-light ratio (1 or 2), we get $1.5\times10^{7}$M$_\odot$ and $3.0\times10^{7}$M$_\odot$. In this case we are thus slightly underestimating the stellar mass, and this is likely to happen because of the small number of old RGB stars present in the shallow CMD.

\subsection{ESO384-16} \label{}

ESO384-16 is considered to be a transition-type or lenticular dwarf \citep[e.g.,][]{jerjen00b, beaulieu06, bouchard07}, with a rather regular elliptical shape and a high central stellar density ($\sim\!3.5$ stars per arcsec$^{2}$ corresponding to $\sim660$ stars per $0.1$ kpc$^{2}$, at luminosities between the TRGB and one mag below the TRGB, and after correcting for incompleteness). The latter characteristic marks the separation between early-type dwarfs and dwarf lenticulars \citep[see also the discussion in][]{beaulieu06}, together with a higher $M_{HI}/L_B$ ratio than the typically close to zero ratio for early-type dwarfs. In the mentioned study, the neutral gas mass is reported to be $\sim\!10^{6}$M$_\odot$, and its spatial distribution is described as asymmetric. For this reason, \citet{beaulieu06} suggest that ESO384-16 is actually falling toward the center of the group. \citet{bouchard07} also suggest that this galaxy may be experiencing a mild ram-pressure stripping from the intra-group medium, based on the asymmetries in its HI distribution. In its CMD, which has $\sim\!17200$ stars, we identify an overdensity of luminous AGB stars above a prominent RGB, but the BL and RSG regions are sparsely populated, indicating very little star formation in the past 100 Myr. However, this galaxy was detected in H$\alpha$ by \citet{bouchard08}. ESO384-16 is located halfway between the dominant giants of the group, at a deprojected distance of $\sim\!1.0\pm0.4$ Mpc from CenA and $\sim\!0.9\pm0.4$ Mpc from M83, and has a negative tidal index \citep{kara07}. According to \citet{georgiev08}, ESO384-16 hosts two globular clusters.

When deriving its SFH, we immediately notice that this galaxy has been one of the most active in our CenA sample, with an average \emph{SFR} of $\sim\!0.006\pm0.0023$M$_\odot$yr$^{-1}$. As reported in Table~\ref{ressfh}, the parameter $b_{14G}$ is the only one being $>1$, meaning a \emph{SFR} higher than the average for ages older than 1 Gyr ago. About $\sim 60\% ^{+15\%}_{-25\%}$ of the stars in ESO384-16 were born more than 5 Gyr ago. The recovered SFH (Fig.~\ref{sfhsnew}) shows two episodes of enhanced star formation, one taking place $\sim\!250-400$ Myr ago, and an older one occurring between 2 and 3 Gyr ago. However, for the latter, old ages, there is a high uncertainty given that the main feature in the CMD is the RGB, clearly degenerate in age and metallicity. In fact, we have to slightly change the initial values of distance and reddening reported in Table~\ref{infogen} to find the best-fitting synthetic CMD for our SFH recovery. The new values we adopt are $E(B-V)=0.12$ (instead of 0.08), and $(m-M)_{0}=28.18$ (instead of $28.28\pm0.14$ found by \citealt{kara07}), which however lie within the errorbars of the previous values. Coming to the most recent star formation in this galaxy, for ages as young as $\sim\!4$ Myr ago our \emph{SFR} lies on the upper limit of the H$\alpha$ measurement given in \citet{bouchard08} (who report a current \emph{SFR} of $\sim\!0.00023\pm0.00006$M$_\odot$yr$^{-1}$). These authors find that the H$\alpha$ emission in ESO384-16 is quite faint and diffuse, thus being in agreement with the evidence from its resolved populations.

As mentioned before, the age-metallicity degeneracy in the RGB phase makes it difficult to constrain the average metallicity for ESO384-16. Considering the relatively high average \emph{SFR} for this galaxy, the best-fit result of [Fe/H]$=-0.97\pm0.15$ is appropriate for its luminosity (related to the total mass). We estimate the total stellar mass to be $8.0^{+2.0}_{-2.5} \times10^{7}$M$_\odot$ from our SFH, while the values derived assuming a stellar mass-to-light ratio of 1 or 2 are $3.0\times10^{7}$M$_\odot$ and $5.6\times10^{7}$M$_\odot$, respectively. The reported values are consistent with each other, if we consider that the number obtained from the SFH is likely to be overestimated and that the information extracted from the CMD is limited for this object.

\subsection{ESO269-58} \label{}

ESO269-58 is perhaps the most intriguing object in this sample, as can already be seen from its rich CMD ($\sim\!136000$ stars). While relatively few BL and RSG stars are present in this galaxy, the most prominent features are a very broad RGB and an extremely well populated luminous AGB region (see Fig.~\ref{fore}). This peculiar irregular galaxy shows a prominent dust lane in its central regions, and contains $\sim\!25\times10^{6}$M$_\odot$ of neutral gas. There exists only one measurement of the H$\alpha$ flux for this galaxy, carried out by \citet{phillips86}. The H$\alpha$ emission is extremely weak. ESO269-58 also has an internal extinction of $A_B\sim\!0.50$, according to LEDA. Its position within the CenA subgroup is rather central, being only $\sim\!300\pm50$ kpc away from CenA and with a tidal index close to 2 \citep{kara07}. Finally, ESO269-58 contains as many as 8 globular clusters \citep{georgiev08}.

The average rate at which ESO269-58 has been forming stars is the highest in our sample ($\sim\!0.07\pm0.04$M$_\odot$yr$^{-1}$). Just as KK196 and ESO384-16, this object had a \emph{SFR} higher than the lifetime average value just for ages $>1$ Gyr ago. ESO269-58 formed $\sim 50\% ^{+15\%}_{-15\%}$ of its stellar content more than 5 Gyr ago. The SFH recovery of this object has been made more difficult by the high crowding exhibited by this object. Its peak stellar density is $\sim\!10$ stars per arcsec$^{2}$ (down to $I\sim\!27$), and the $1\sigma$ photometric errors at a magnitude of $I=25$ are already 0.15 mag in magnitude and 0.24 mag in color, which has the effect of significantly broadening the observed features in the CMD. The observed broadening could in part also be due to differential reddening, but the fact that the TRGB is not stretched along the reddening vector, and that the MS does not look heavily affected, points toward ordinary crowding effects. We find new best-fitting values for the reddening and the distance modulus, namely $E(B-V)=0.15$ instead of 0.10, and $(m-M)_{0}=27.80$ instead of $27.90\pm0.16$. As can be seen from the SFH plotted in Fig.~\ref{sfhsnew}, ESO269-58 experienced an enhanced star formation between $\sim\!3$ and 5 Gyr ago, which was followed by a rapid chemical enrichment. This is confirmed by the curvature of the RGB and the metal-rich red turnover of the luminous AGB (see Fig.~\ref{fore}), and the resulting mean metallicity ([Fe/H]$=-0.98\pm0.20$) is in agreement with the absolute luminosity of this galaxy. A second, mild enhancement occurred $\sim\!300$ Myr ago, but overall the activity of this galaxy has been much lower than the average \emph{SFR} in the last 1 Gyr (the fraction of stars born in this period is only $1\%$, see Table~\ref{infogen}). The H$\alpha$ flux reported by \citet{phillips86} is compared to the most recent bin of our SFH (past $\sim8$ Myr), but is slightly smaller than our derived value.

\begin{figure*}
 \centering
  \includegraphics[width=17.cm]{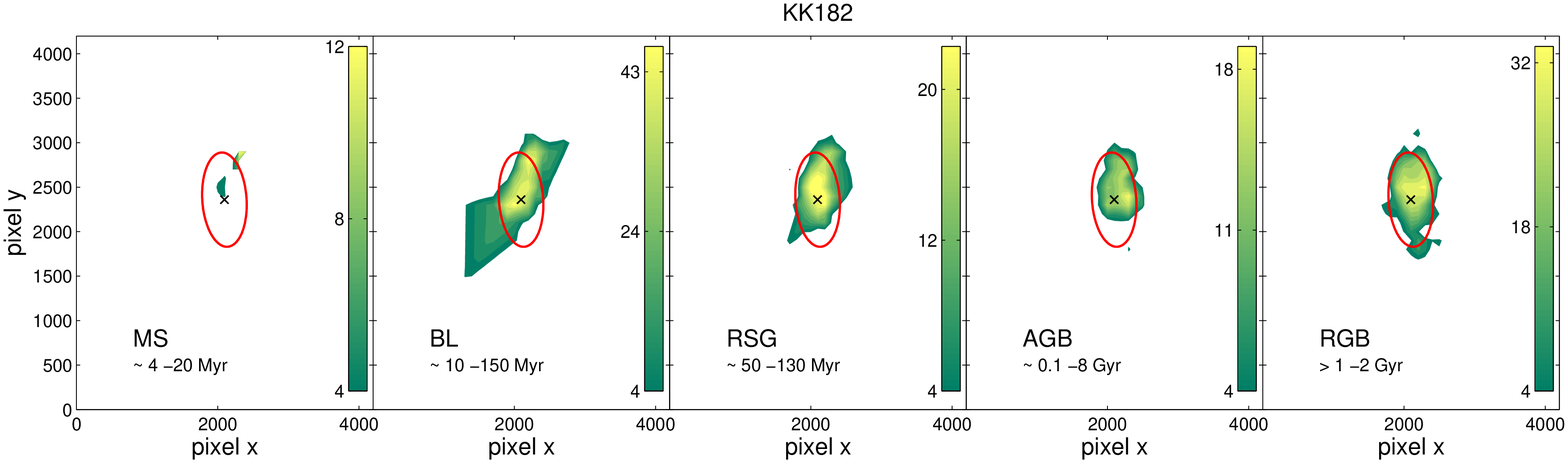}
  \includegraphics[width=17.cm]{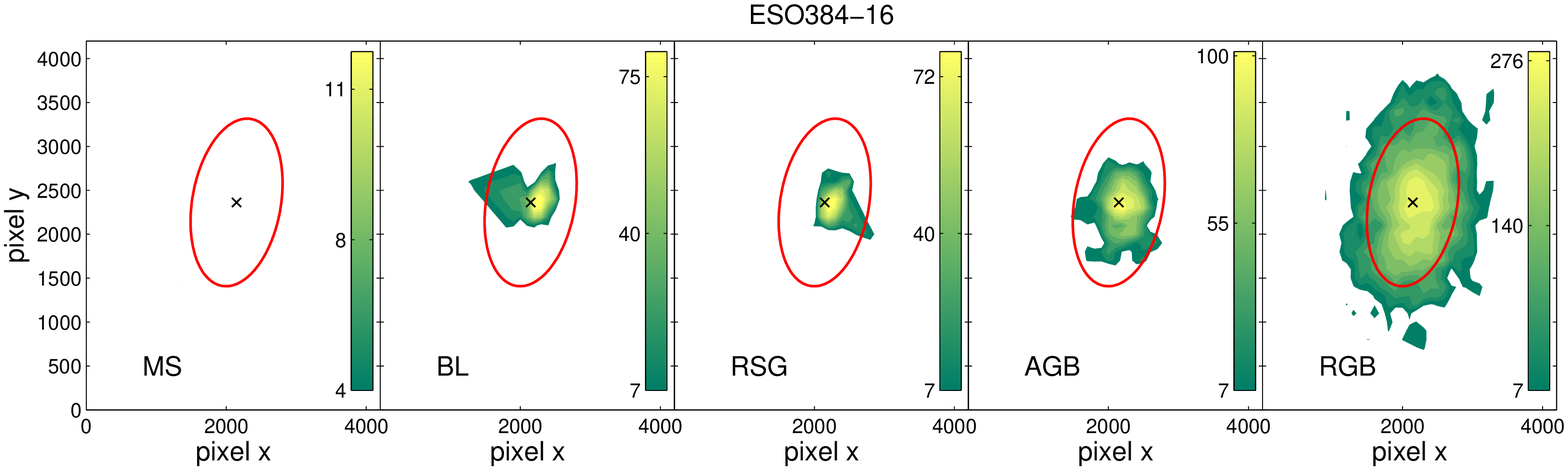}
  \includegraphics[width=17.cm]{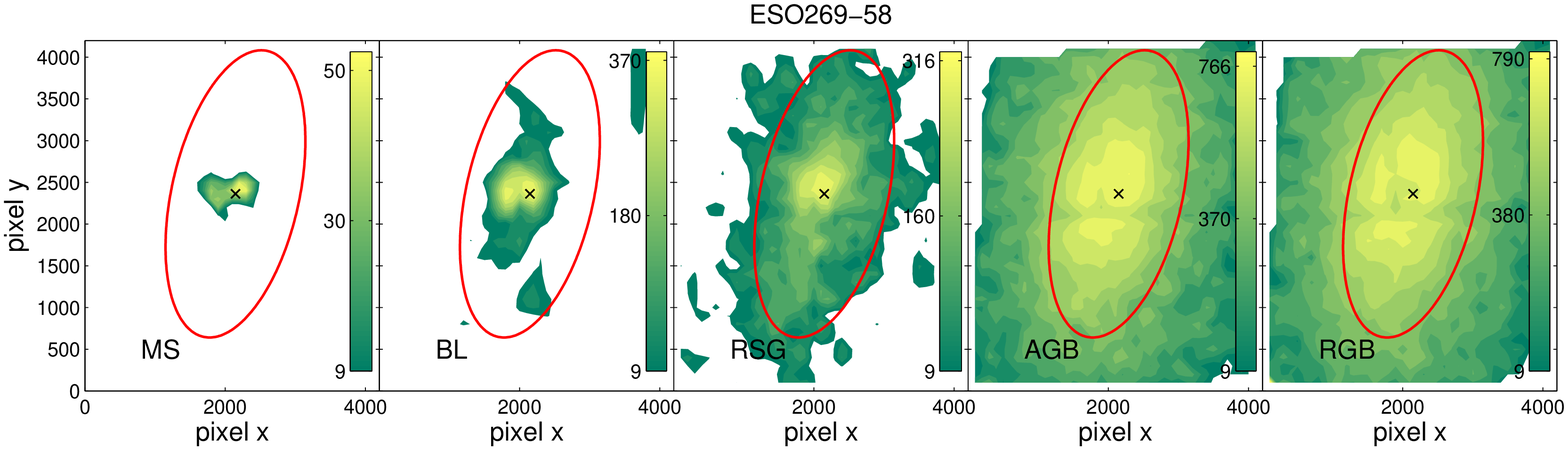}
 \caption{\footnotesize{Density maps for three of our target galaxies (KK182, ESO384-016, and ESO269-58, ordered by absolute magnitude), each divided in five stellar evolutionary phases. These are: MS, BL, RSG, luminous AGB and RGB, ordered by increasing age. Note that the ages corresponding to each stage (reported in the top panel) are just estimates. The stellar density values are listed along the colorbars, in units of number of stars per $0.1$ kpc$^{2}$. 10 equally spaced isodensity contours are drawn starting at the $1\sigma$ significance level up to the peak significance level. The peak levels are: for KK182: MS=$2\sigma$, BL=$3.4\sigma$, RSG=$2.6\sigma$, AGB=$2.5\sigma$, RGB=$3.1\sigma$; for ESO384-16: MS=$1.5\sigma$, BL=$3.4\sigma$, RSG=$3.4\sigma$, AGB=$3.7\sigma$, RGB=$4.7\sigma$; for ESO269-58: MS=$2.7\sigma$, BL=$4.7\sigma$, RSG=$4.5\sigma$, AGB=$5.4\sigma$, RGB=$5.5\sigma$. The center of each galaxy is indicated with a black cross. Just as a reference among different frames, we also overplot in red the ellipse corresponding to the projected major axis radius at the isophote level of 25 mag arcsec$^{-1}$ in the $I$-band \citep[taken from ][]{sharina08}.}}
 \label{denmap2}
\end{figure*}

\citet{davidge07} also studied the resolved stellar populations of ESO269-58 in its outskirts, using ground-based optical data. He reports a mean metallicity value of [Fe/H]$\sim\!-1.8$, derived by comparing the RGB to isochrones. He also concludes that ESO269-58 harbors a small intermediate-age population (with stars not younger than 1 Gyr), which accounts for a few percent of the galaxy's stellar mass. According to \citet{davidge07}, these stars are the result of an episode of enhanced star formation no more than $\sim\!1$ Gyr ago. The reason for the discrepancy between his study's metallicity value and our results could be due to the blue extension of the RGB. \citet{davidge07} does not resolve stars younger than 1 Gyr, while our CMD shows such populations (the depth of the CMDs are comparable). We thus hypothesize that what he treats as blue, and thus metal-poor, RGB stars in the computation of photometric metallicities, are actually younger, RSG stars blended with the blue part of the RGB due to photometric errors (see also isochrones in Fig.~\ref{isos}). This may lead to an artificially lower mean metallicity as derived from RGB stars. To test our hypothesis, we look at the stellar spatial distributions of a sample of stars along the blue edge of the RGB ($0.5\lesssim V-I \lesssim0.9$) and of a second sample of stars along the mean locus of the RGB ($1\lesssim V-I \lesssim2$), at a magnitude of $I\sim\!25$. The latter are clearly distributed across the whole spatial extent of the galaxy, while the bluer stars are more centrally concentrated, thus indicative of a younger RSG population. Finally, this is supported by the CMDs in Figs.~2 and 3 of \citet{davidge07}, which show relatively more stars on the blue side of the RGB at radii $<4$ kpc than those at larger radii.

We compute the total stellar mass stemming from the recovered SFH, which results in a value of $8.8^{+1.1}_{-2.3} \times10^{8}$M$_\odot$. Note that this galaxy clearly extends further out than the ACS FoV (see Fig. \ref{denmap2}), and we are thus missing part of its stellar content in our mass estimate. On the other hand, when we assume a stellar mass-to-light ratio of 1 or 2, the expected stellar mass values are between $1.0\times10^{8}$M$_\odot$ and $2.1\times10^{8}$M$_\odot$. In this case, where a high number of stars was produced at ages $>5$ Gyr ago, perhaps a higher value of the stellar mass-to-light ratio could be more appropriate \citep[$3-4$, see e.g.][]{cote00}, thus reconciling the two total mass values.


\section{Spatial distribution of stellar populations as a function of time} \label{maps_sec}

It is an established fact that dwarf irregular galaxies host one to several active star forming regions (depending on the host's mass) which are distributed in clumps throughout the galaxy's body \citep[e.g.,][and references therein]{grebel04a}. As already done in \citet{crnojevic11b}, we now have a look at the spatial distribution of stellar populations with different ages in our galaxies. Fig. \ref{denmap2} shows the density maps for three of the target galaxies' stellar age subsamples mentioned in Sect. \ref{cmd_sec} (see Fig. \ref{isos}). To separate the galaxies' stellar content into MS, BL, RSG, AGB and RGB stars (in order of increasing age), we only select regions in the CMD that have photometric errors smaller than $\sim0.1$ mag in magnitude and $\sim0.15$ mag in color to avoid as much as possible the overlap of different evolutionary stages when the errors are larger (e.g., MS and BL stars). For HIPASS J1348-37 and KK196, the number of stars in the MS, BL and RSG stages is too small to draw density maps, so we do not plot them for the mentioned objects. Also ESO384-16 lacks a significant MS population for which we can derive density maps (see Fig.~\ref{denmap2}).

The density maps are computed by counting for each star the number of neighbours found within $\sim\!0.03-0.07$ kpc$^{2}$, depending on the distance of the galaxy (arbitrary value chosen such that we do not add too much substructure but we still retain the overall features). Note that at distances of $\sim4$ to 6 Mpc, 1 arcsec corresponds to $\sim\!0.02-0.03$ kpc. The result is then convolved with a square grid, with a final resolution of $0.01-0.02$ kpc$^{2}$. On the density map, we draw 10 equally spaced isodensity contours, starting from a $1\sigma$ significance level (corresponding to $\sim4$ stars per $0.1$ kpc$^{2}$) up to the peak significance level, different for each map and indicated in the caption of Fig. \ref{denmap2}. The stellar density for each subsample is then reported in the colorbars (units of stars per $0.1$ kpc$^{2}$).

In the top panel of Fig. \ref{denmap2} we also report the age range covered by each evolutionary stage in our CMDs, although these are just estimates which do not take into account the spread of stars due to photometric errors. The youngest populations (a few Myr up to $\sim100$ Myr) in dwarf irregular galaxies are generally found in concentrated, actively star forming regions, which are located in the center as well as in offset knots \citep[see, e.g.,][]{vandyk98, glatt10}. This is clearly observed in the MS and/or BL/RSG density maps of our targets, where the relative star formation is confined to the dwarf's center. Active star forming regions tend to turn on and off in adjacent cells with time in a stochastic propagation fashion, with star forming complexes having lifetimes on the order of $10^2$ Myr and sizes of up to some hundreds of pc in diameter \citep{seiden79, dohm97, grebel98, dohm02, weisz08, glatt10, crnojevic11b}. Interestingly, the triangular shape of KK182 comes from an offset knot of star formation which is clearly recognized as a local enhancement in its MS and BL density maps, while it fades away in the RSG map. Note that while BL stars with ages $>150$ Myr are present in our targets, they are left out from our selection box because of the adopted faint magnitude cut (used to avoid CMD regions with too large photometric errors, see above). The RSG selection box, on the other hand, contains stars with ages of $\sim50$ up to 130 Myr (see Fig.~\ref{isos}). The BL and RSG evolutionary phases present on average smoother and more extended density maps than the MS, due to the fact that stars with these ages tend to migrate away from their original birthplace thus erasing the initial substructures \citep[e.g.,][and references therein]{bastian11}. Finally, what is observed for older stars (luminous AGB and RGB, with ages $\gtrsim0.5$ Gyr) is a broadened elliptical distribution encompassing the entire galaxy, which represents the redistribution of the stars over long timescales (see also \citealt{zaritsky00}; \citealt{battinelli07}; Fig. 9 in \citealt{crnojevic11b}).


\section{Discussion} \label{discuss}

\begin{figure}
 \centering
  {\includegraphics[width=9cm]{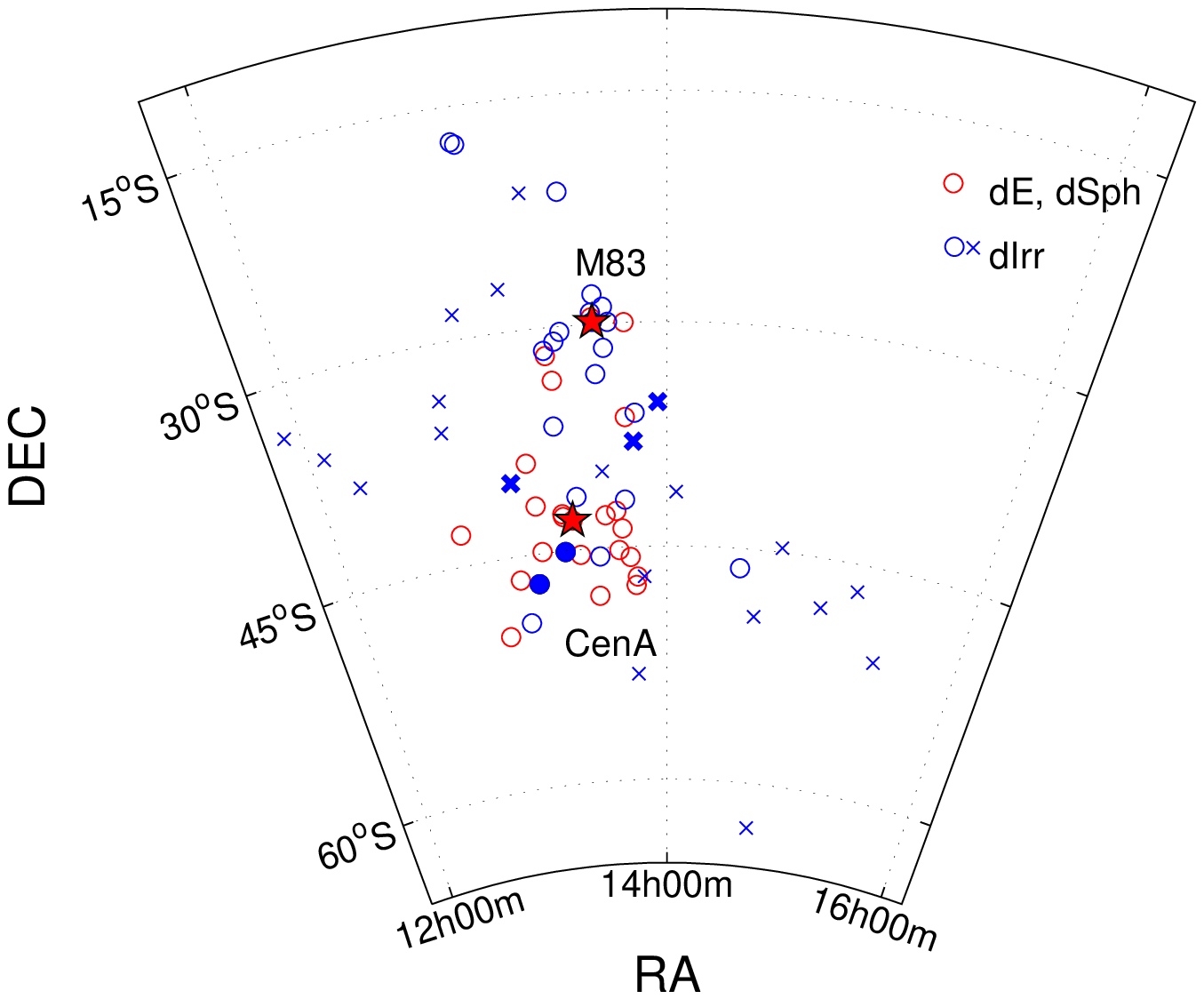}
  \includegraphics[width=9cm]{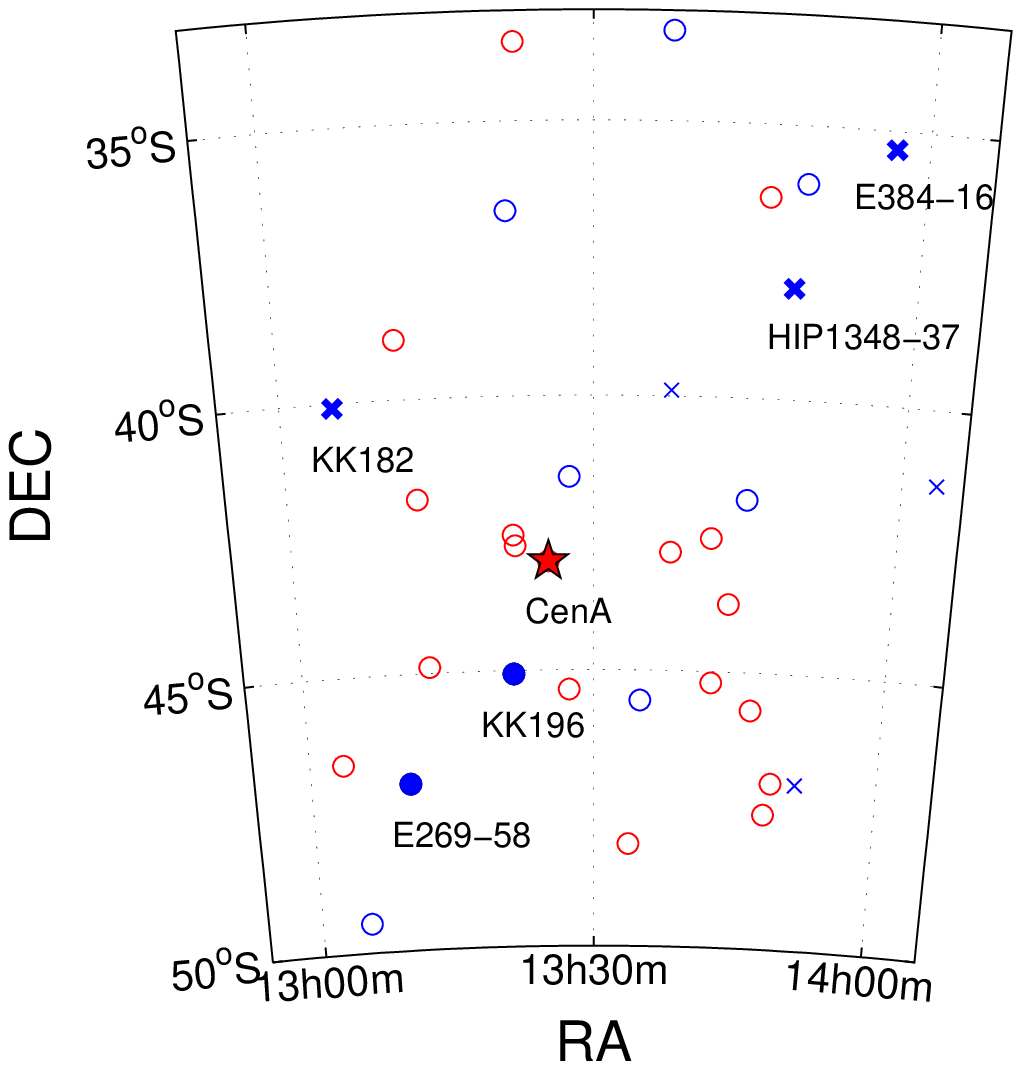}}
 \caption{\footnotesize{\emph{Upper panel.} Positions in the sky of the galaxies belonging to the Centaurus A/M83 complex (from the coordinates given in \citealt{kara07}). Red symbols indicate early-type dwarfs (dwarf ellipticals, dEs, and dwarf spheroidals, dSphs), while blue symbols are for late-type dwarfs (dIrrs). The circles are objects with positive tidal index, while the crosses objects with negative tidal index (i.e., isolated objects). Two red stars indicate the positions of the two dominant giant galaxies (the elliptical CenA and the spiral M83), around which the smaller companions are clustering, forming two distinct subgroups. The filled circles and thick crosses represent the dwarfs studied here. \emph{Lower panel.} Same as above, this time zoomed-in on a smaller region around CenA, where our sample of dIrr galaxies is located (as labeled in the plot).}}
 \label{sky}
\end{figure}

The galaxies in our sample span a considerable range in their physical characteristics. Their luminosities vary from $M_B=-11.90$ to $-14.60$, their neutral gas content ranges from undetectable to few times $10^7$M$_{\odot}$, some of them are located in high-density regions (in terms of number of galaxies) and others are isolated objects. Together with the M83 subgroup dwarfs studied in \citet{crnojevic11b}, we have a very heterogeneous sample of galaxies, and their SFHs appear to be similarly diverse.

We have found that for the target galaxies the star formation initially proceeds at about the lifetime average rate (at ages $\gtrsim5$ Gyr). We note that, however, we do not have the resolution to detect possible bursts of star formation at old ages. In the case of KK182, a very pronounced activity is found in the last Gyr, with brief and fluctuating star formation episodes that reach in intensity more than five times the lifetime average \emph{SFR}. On the other hand, for HIPASS J1348-37 and KK196 the \emph{SFR} remains at about the average value in the first Gyrs, subsequently proceeds with mild amplitude variations, and almost disappears in the last $\sim\!100$ Myr. Given that there is an H$\alpha$ detection for KK196 and no HI gas appears to be present, out of these two objects only HIPASS J1348-37 is a candidate transition-type dwarf. However, KK196 holds a peculiar position within the group, being possibly affected by CenA's radio lobe. In the cases of ESO384-16 and ESO269-58, the \emph{SFR} was more pronounced between $\sim\!2$ and $\sim\!3-4$ Gyr ago, while at more recent times the \emph{SFR} was well below the average value but showed a single peak a few hundred Myr ago. ESO384-16 is indeed classified as a transition-type dwarf (see the relevant subsection). Both galaxies contain several globular clusters, although like ESO444-78 in the M83 sample, the neutral gas content and the luminosities are larger than those of comparable Local Group transition-type dwarfs. 

Overall, the studied objects seem to follow the expected regime of a ``gasping'' star formation, where quiescent periods with low \emph{SFR} are interrupted by periods of high activity (with respect to the average lifetime), rather than having a constant level of star formation activity throughout their lives \citep[e.g.,][]{marconi95, tolstoy09}. Moreover, in agreement with previous studies we find the most recent (i.e., a few hundreds Myr) star formation in our targets to be generally concentrated in clumps around the central region of the galaxies \citep[see, e.g.,][]{harbeck01}. ESO269-58 stands out of our sample as the object with the highest lifetime average \emph{SFR}, $\sim\!6 \times 10^{-2}$M$_\odot$yr$^{-1}$ with respect to the few times $10^{-3}$M$_\odot$yr$^{-1}$ of the other dwarfs in both the CenA and M83 subgroups. Just as observed in the Local Group, we see a variety of SFHs. In the LG we have cases where most of a dwarfs' stellar content was formed at ancient times \citep[$\gtrsim10$ Gyr, e.g.,][]{grebel04}, but also cases where the bulk of star formation occurred at ages younger than that (e.g., IC1613 or Leo A, see \citealt{skillman03, cole07}), and this appears to be true also for the galaxies in our sample. We will come back to this point later.

The CenA/M83 group is shown projected on the sky in the upper panel of Fig. \ref{sky}, with the five galaxies studied here highlighted in the lower panel of the same figure, a more detailed view of the CenA subgroup. As can be easily noticed, the latter has a much higher number of early-type dwarfs, which reflects a probably more advanced evolutionary stage as compared to the dwarf-irregular-rich environment of M83. In fact, in our sample HIPASS J1348-37, KK182, and ESO384-16 all are rather isolated objects, being no closer than 900 kpc from either of the group central giants. We can thus compare the derived properties of dIrrs that are bound members of the group to those that are at its edges. There is no obvious indication of environmental effects on the evolution of our dwarfs from looking at the shape of their SFHs alone. In fact, we can see that galaxies that have rather similar shapes of their SFH (e.g., HIPASS J1348-37 and KK196, as well as ESO384-16 and ESO269-58) are located in very different environments, being either close to the central giant of the group or at its outskirts. However, we must bear in mind that we only know present-day distances for our targets.

\begin{figure}
 \centering
  {\includegraphics[width=8.5cm]{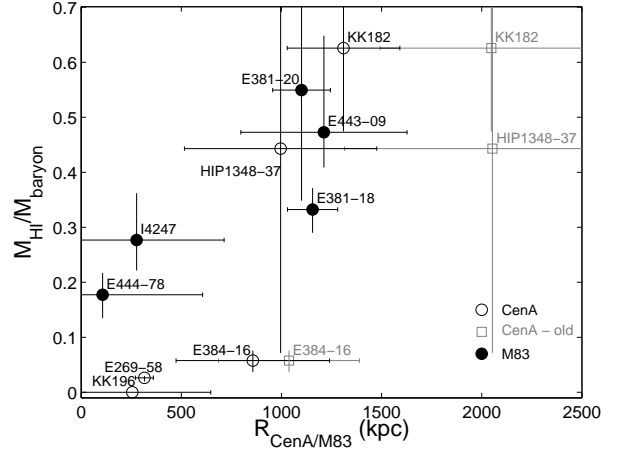}}
\caption{\footnotesize{Ratio of neutral gas mass to baryonic mass ($=M_{star}+M_{gas}$) as a function of deprojected distance from the dominant subgroup galaxy. The black open circles represent the CenA subgroup late-type dwarfs, while the black filled circles are M83 subgroup members (data taken from \citealt{crnojevic11b}). Grey squares indicate the old distances for three of the CenA subgroup members, while black open circles for the same galaxies are new distances that we recompute because they are actually closer to M83 than to CenA (see text for details).}}
 \label{mhi_vs_dist}
\end{figure}

Three of the studied galaxies are metal-poor ([Fe/H]$\sim\!-1.4$ dex), while two of the most luminous galaxies in our sample, ESO384-16 and ESO269-58, have slightly higher metallicities ([Fe/H]$\sim\!-1.0$ dex) with respect to the others, thus showing a hint of a metallicity-luminosity relation. Unfortunately, it is difficult to compare these results to those previously obtained for the sample of early-type dwarf companions of CenA \citep{crnojevic10}. This is because the estimate of the metallicity for the late-type dwarfs is an average that includes both old and young stars (SFH recovery method), while the metallicity values for early-type dwarfs come primarily from their old populations (photometric metallicity estimation). Moreover, we used the Dartmouth stellar evolutionary models \citep{dotter08} to derive photometric metallicities in our previous study, while for the present SFH recovery we adopted the Padova set of models (since only these models cover the full range of ages that we need). The use of these two models gives metallicity results that differ by up to approximately 0.3 dex (judging from the upper part of the RGB).

\subsection{Neutral gas content}

The neutral gas masses of the target dwarfs do not correlate with their present-day distance from CenA/M83. Given that we know the HI masses of our targets (Tab. \ref{infogen}) and that we inferred their stellar masses from the derived SFHs (Sect. \ref{sfh_sec}), we can easily compute their baryonic masses by adding these quantities and multiplying the gas mass by 1.33 in order to account for the He content \citep{matthews98}. We then plot the ratio of neutral gas mass to baryonic mass as a function of deprojected distance from the dominant subgroup galaxy (Fig.~\ref{mhi_vs_dist}). When considering the ratio of neutral gas to total baryonic mass there is a clear correlation between the two quantities, that could be explained with an external culprit depriving the dwarfs in the central part of the group from their gas content, and as a consequence also from the possibility of forming new stars \citep[see also the discussion in][]{weisz11}. This is not a firm conclusion, however, because we lack information on the space motions of the dwarfs that would tell us how their distance from the dominant galaxy has changed over their lifetime.

In addition, as for the M83 dwarf companions that are found in the very center of their respective subgroup, also ESO269-58 ($\sim\!320$ kpc from CenA) would need only $\sim\!350$ Myr at its average \emph{SFR} to exhaust its remaining neutral gas content, while in the other CenA companion with positive tidal index, KK196, no HI gas has been detected so far. On the contrary, the galaxies of our sample that currently have a negative tidal index show instead high $M_{HI}/\!<SFR>$ ratios ($\geq10^{10}$), which is in qualitative agreement with the results of \citet{bouchard08}. The exception is ESO384-16, which has a $M_{HI}/\!<SFR>$ ratio of $850$ Myr but, as mentioned before, this object could possibly be experiencing ram-pressure stripping \citep{bouchard07}. Coupled with the information that the amount of neutral gas per unit baryonic mass is lower for galaxies closer to the center of the group, a shorter gas consumption timescale is possibly linked to an external effect (see discussion below) rather than to properties intrinsic to the dwarfs.

\subsection{Star formation parameters as a function of environment}

\begin{figure}
 \centering
  {\includegraphics[width=8.5cm]{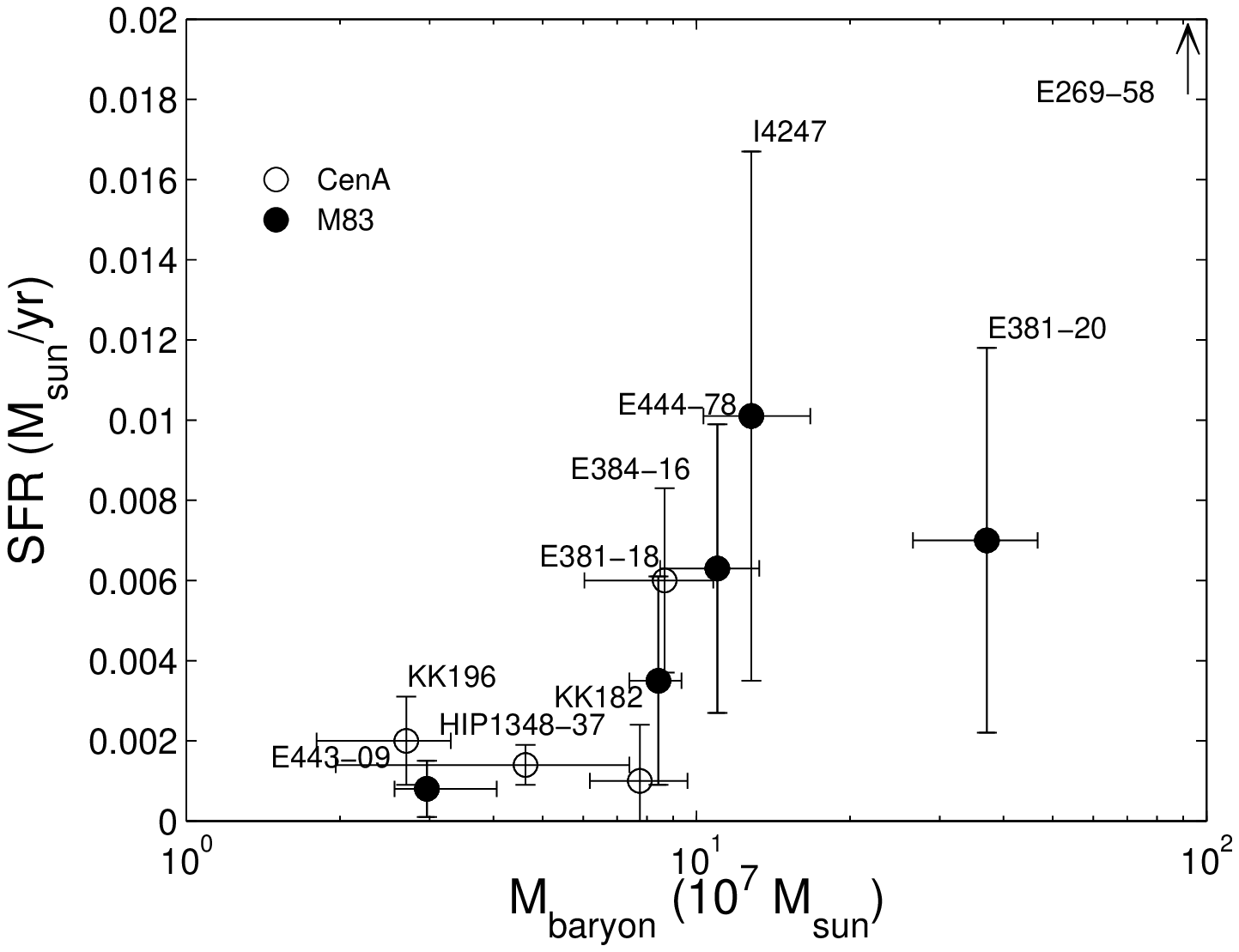}
  \includegraphics[width=8.5cm]{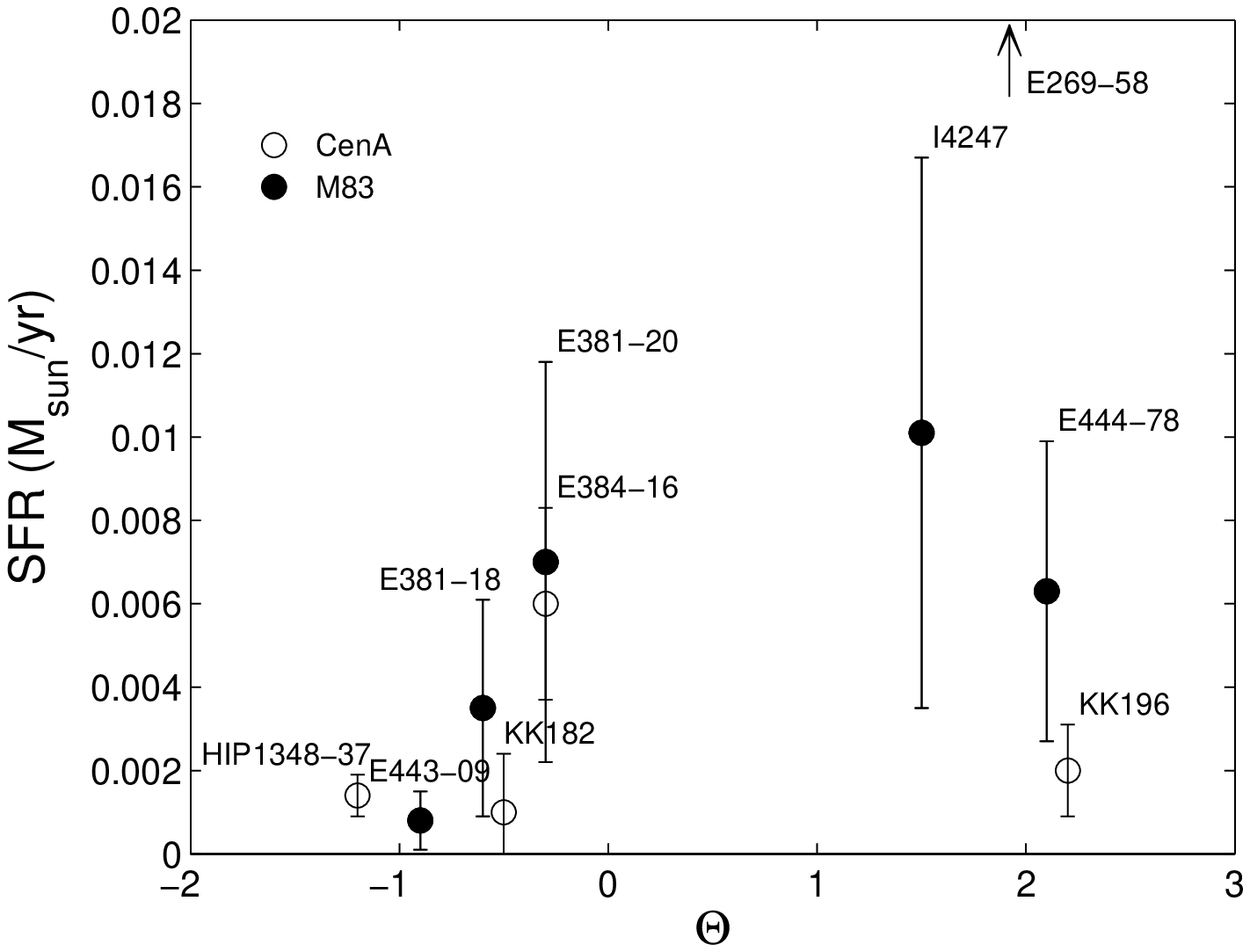}
  \includegraphics[width=8.5cm]{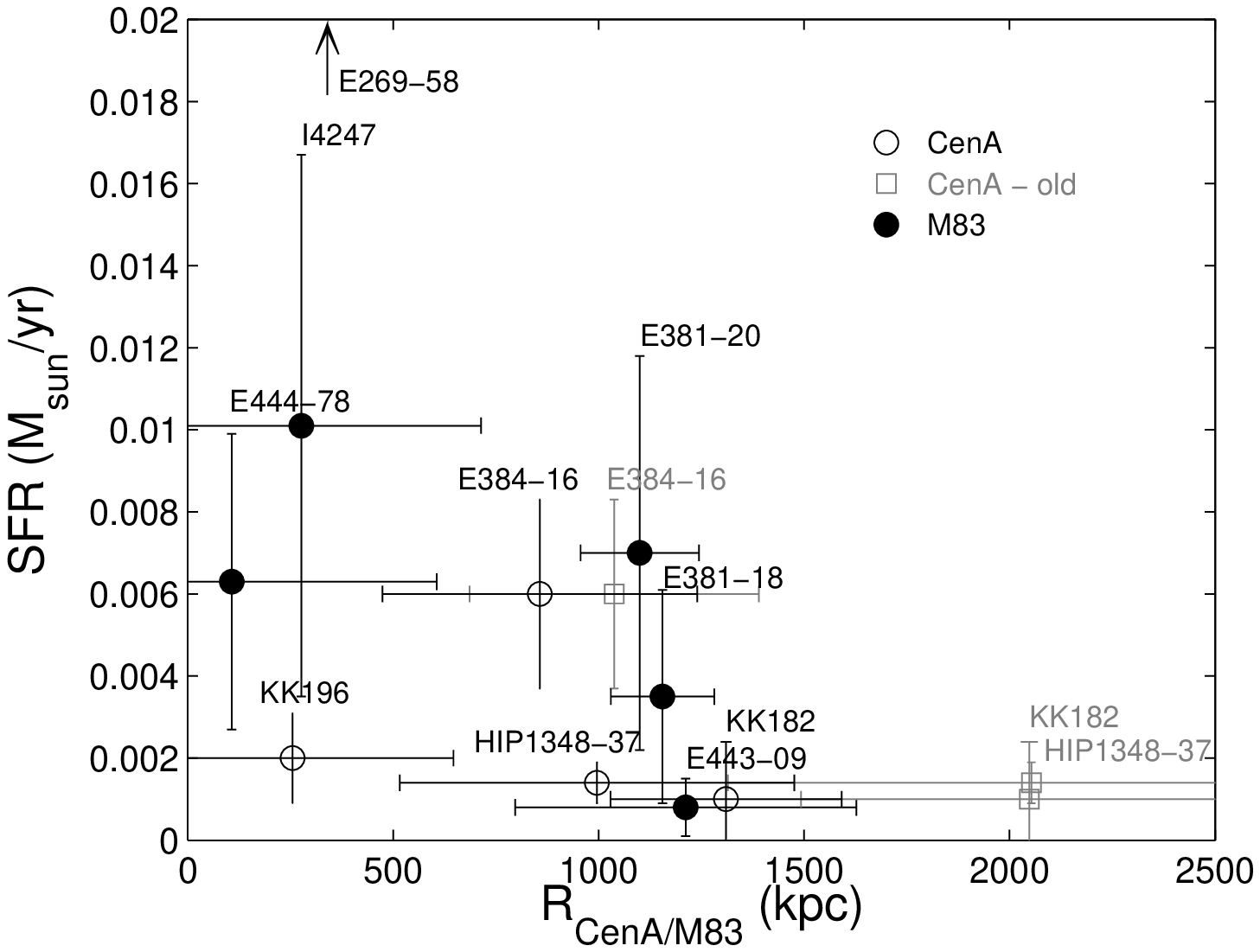}}
\caption{\footnotesize{Lifetime average star formation rate plotted as a function of (\emph{top panel}) baryonic mass ($=M_{star}+M_{gas}$), (\emph{central panel}) tidal index, and (\emph{bottom panel}) deprojected distance from the dominant subgroup galaxy. The black open circles represent the CenA subgroup late-type dwarfs, while the black filled circles are M83 subgroup members (data taken from \citealt{crnojevic11b}). 
In the bottom panel, grey squares indicate the old distances for three of the CenA subgroup members, while black open circles for the same galaxies are new distances that we recompute because they are actually closer to M83 than to CenA (see text for details). For ESO269-58, the \emph{SFR} is much higher than those for the other objects ($\sim\!0.07\pm0.04$M$_\odot$yr$^{-1}$), so this value falls outside our plot limits. In the panels, we only indicate its luminosity/tidal index/deprojected distance values with an arrow.}}
 \label{sfr_vs_lt}
\end{figure}

\begin{figure}
 \centering
  {\includegraphics[width=9cm]{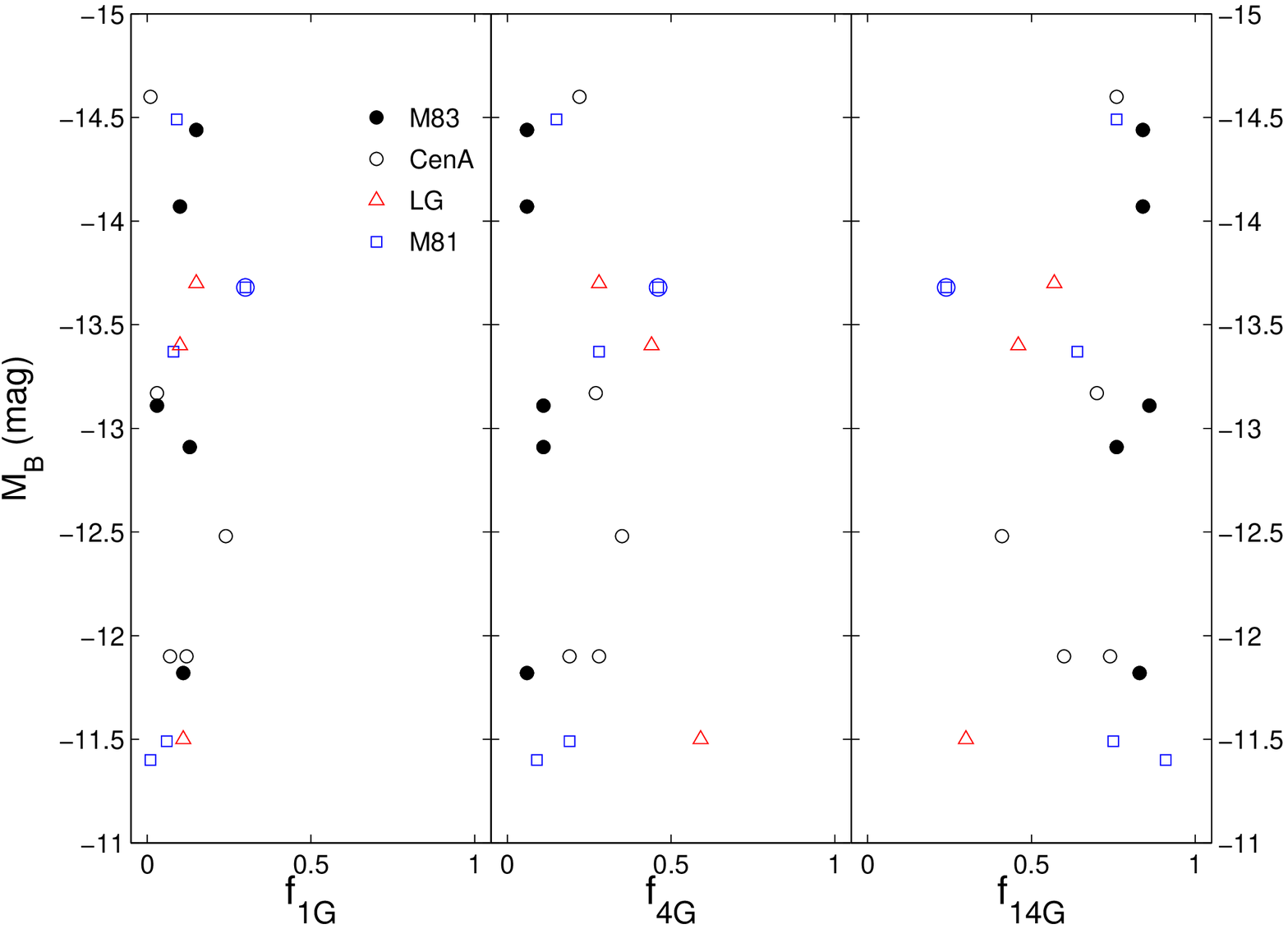}
  \includegraphics[width=9cm]{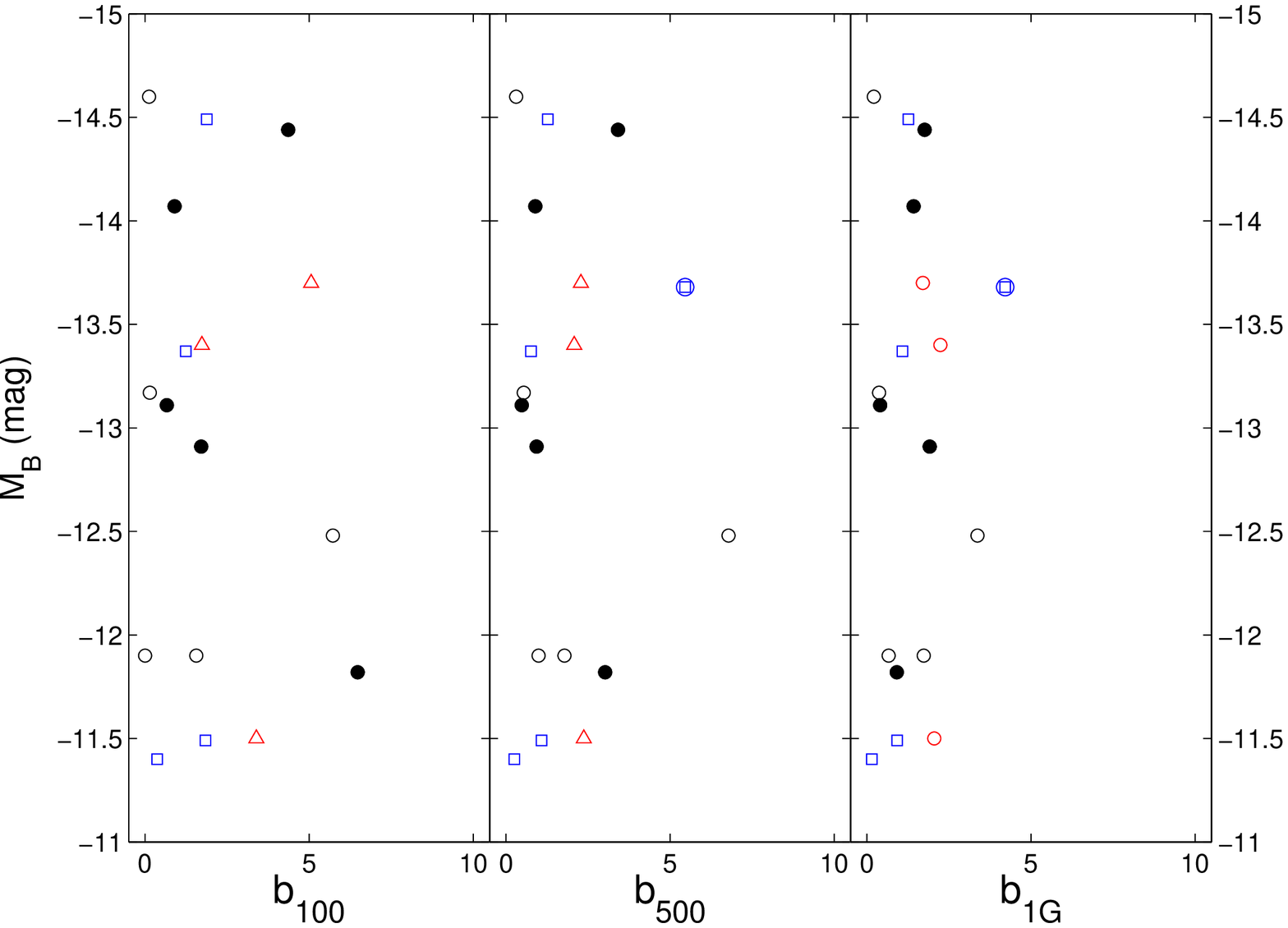}}
\caption{\footnotesize{Absolute $B$-band luminosity plotted as a function of: (\emph{upper panel}) the fraction of stars formed in the past $0-1$ Gyr ($f_{1G}$), $1-4$ Gyr ($f_{4G}$) and $4-14$ Gyr ($f_{14G}$); (\emph{lower panel}) the ratio of \emph{SFR} to the lifetime average \emph{SFR} over the last 100 Myr ($b_{100}$), 500 Myr ($b_{500}$) and 1 Gyr ($b_{1G}$). The black open circles represent the CenA subgroup late-type dwarfs, the black filled circles are M83 subgroup members, red triangles indicate Local Group late-type dwarf members and blue squares the M81 group late-type dwarfs. Holmberg IX is a candidate tidal dwarf in the M81 subgroup, and we mark it with an additional larger, open blue circle since this object is not a typical dwarf galaxy (see text for details).}}
 \label{lum_vs_lt}
\end{figure}

As anticipated in \citet{crnojevic11b}, we now want to investigate possible trends of the star formation parameters derived in Sect. \ref{sfh_sec} (Tab. \ref{ressfh}) with environment.

We thus analyze the specific properties of our targets as a function of different parameters, i.e., luminosity/baryonic mass, tidal index, and deprojected distance from the closest giant galaxy. First of all, given that to first order light traces the baryonic mass, we find the average \emph{SFR} to be increasing with luminosity \citep[see, e.g.,][and references therein]{grebel04a}. We then show the average \emph{SFR} as a function of baryonic mass in the top panel of Fig.~\ref{sfr_vs_lt}, and confirm the correlation between these two quantities, thus highlighting the importance of intrinsic properties in the evolution of dwarfs. There is no clear correlation between the average \emph{SFR} and the tidal index of our galaxies (central panel of the same figure), and the same is also true for the deprojected distance (bottom panel). For the latter, we plot the distances based on the original group classification, and then also add the recomputed data points for the three CenA companions that are actually closer to M83 (HIPASS J1348-37, KK182, and ESO384-16). If we consider these three objects as part of the CenA subgroup, it could be possible to claim that a mild trend is present, where the \emph{SFR} is decreasing with increasing distance from the dominant galaxy. However, after recomputing their distances from M83, no clear correlation is seen between \emph{SFR} and distance, only a scatter in the values of the sample. The absence of a trend seems reasonable if we bear in mind that the positions of the galaxies in the group have changed with time, and so the present-day position is not necessarily indicative of the distance from other galaxies in the past. We have no way of reconstructing the orbits of our sample galaxies within the group. In these plots, ESO269-58 appears to be an outlier given its high average \emph{SFR}, and this result still holds if we use the \emph{SFR} to $L_B$ ratio instead of the \emph{SFR} alone. One of the possible explanations could be that ESO269-58 is on its first infall into the group \citep[e.g.][]{pasetto11}. Alternatively, the absence of a trend with distance might reinforce the role that intrinsic properties play in the evolution of the target dwarf galaxies.

We also computed parameters to quantify the amount of star formation in certain time periods, relative to the average value, and the fraction of stars born at young, intermediate-age and old epochs (Tab.~\ref{ressfh}). We now want to investigate how these parameters behave as a function of luminosity, tidal index and deprojected distance from CenA/M83. A similar study was done by \citet{weisz08}, concentrating on M81 and Local Group dwarf irregulars with a larger luminosity range than the one considered here. Since the chosen filters, the depth of the CMDs and the adopted techniques are similar, their sample can be compared to ours. \citet{weisz08} do not find any clear trend of the luminosity with any of the parameters, and the values for the interacting M81 group and the Local Group do not reveal any substantial difference. They conclude that, given that the average \emph{SFR} within their galaxy sample is almost consistent with a constant \emph{SFR} over a Hubble time, the observed intrinsic scatter in the parameters may indicate the stochastic nature of the star formation process for these objects. We consider the galaxies from their sample (both from the M81 and the Local Group) that are in the same luminosity range as those in our sample, and investigate the cited parameters as a function of luminosity. Some of our results are very similar to theirs. For example, we do not find any correlation between absolute $B$ luminosity and $f$ (fraction of stars formed in certain time periods) or $b$ (ratio of the \emph{SFR} in a certain time period to the lifetime average \emph{SFR}) parameters, which are plotted in Fig.~\ref{lum_vs_lt}. We show the values for dwarf galaxies belonging to three different groups (CenA/M83 group, Local Group and M81 group), and we do not see a clear trend in their values. We note that one of the M81 group members (Holmberg IX) is classified as a candidate tidal dwarf \citep[e.g.,][]{makarova02}. This type of dwarf is supposed to have formed out of the baryonic material coming from a close interaction episode of the parent galaxy (in this case, the spiral M81), and thus a tidal dwarf should be a dark matter-free object \citep[e.g.,][]{toomre72, duc97, duc00}. One of the peculiarities of tidal dwarfs is that they contain very few old populations (most probably coming from the parent galaxy), and a more pronounced young population, born after the interaction event. They will thus not share the common properties of a typical dwarf galaxy, and we decide to mark Holmberg IX as an outlier in the relations discussed here.

\begin{figure}
 \centering
  {\includegraphics[width=9cm]{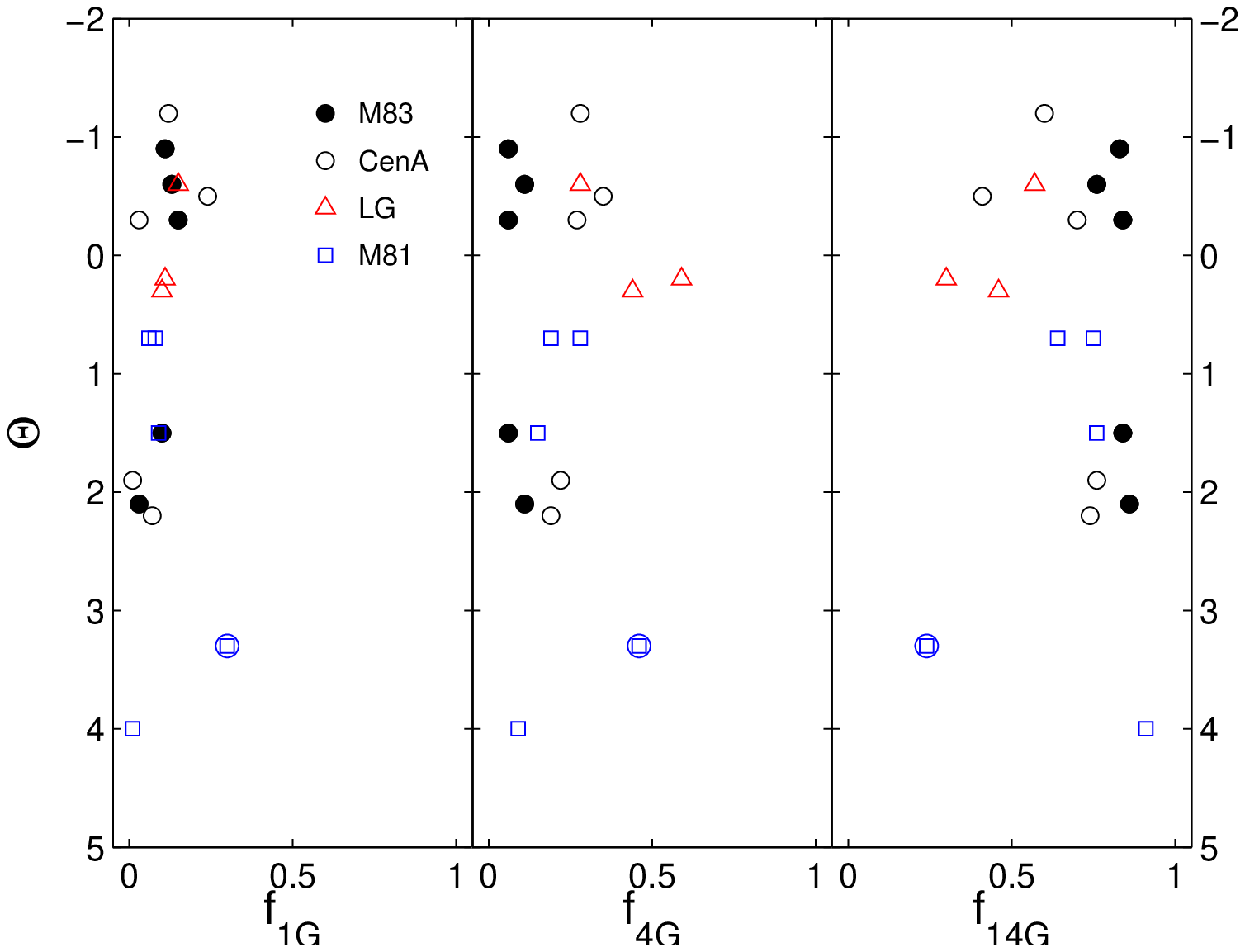}
  \includegraphics[width=9cm]{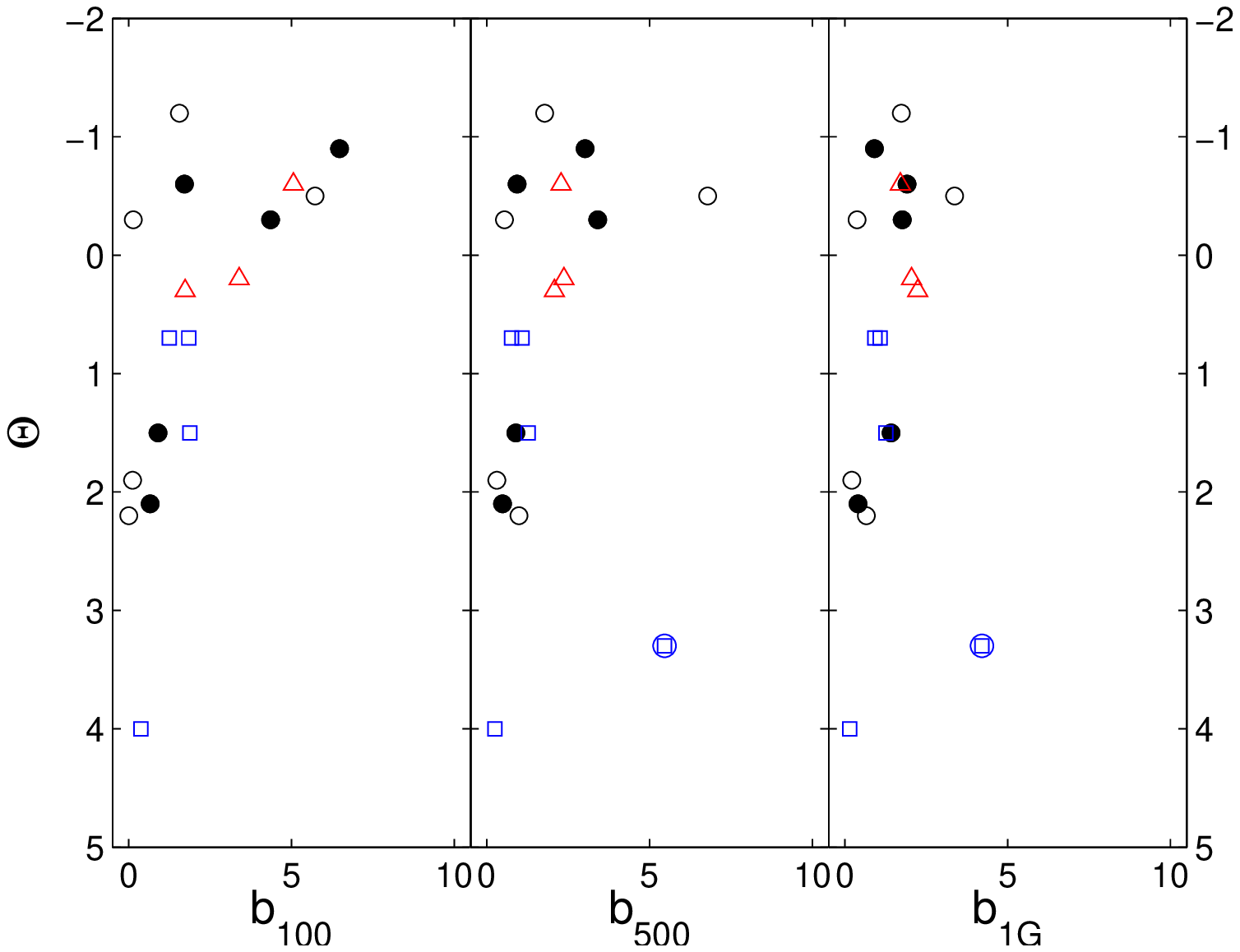}}
\caption{\footnotesize{Tidal index plotted as a function of: (\emph{upper panel}) the fraction of stars formed in the past $0-1$ Gyr ($f_{1G}$), $1-4$ Gyr ($f_{4G}$) and $4-14$ Gyr ($f_{14G}$); (\emph{lower panel}) the ratio of \emph{SFR} to the lifetime average \emph{SFR} over the last 100 Myr ($b_{100}$), 500 Myr ($b_{500}$) and 1 Gyr ($b_{1G}$). The black open circles represent the CenA subgroup late-type dwarfs, the black filled circles are M83 subgroup members, red triangles indicate Local Group late-type dwarf members and blue squares the M81 group late-type dwarfs. Holmberg IX is a candidate tidal dwarf in the M81 subgroup, and we mark it with an additional larger, open blue circle since this object is not a typical dwarf galaxy (see text for details).}}
 \label{tid_vs_lt}
\end{figure}

\begin{figure*}
 \centering
  {\includegraphics[width=11cm]{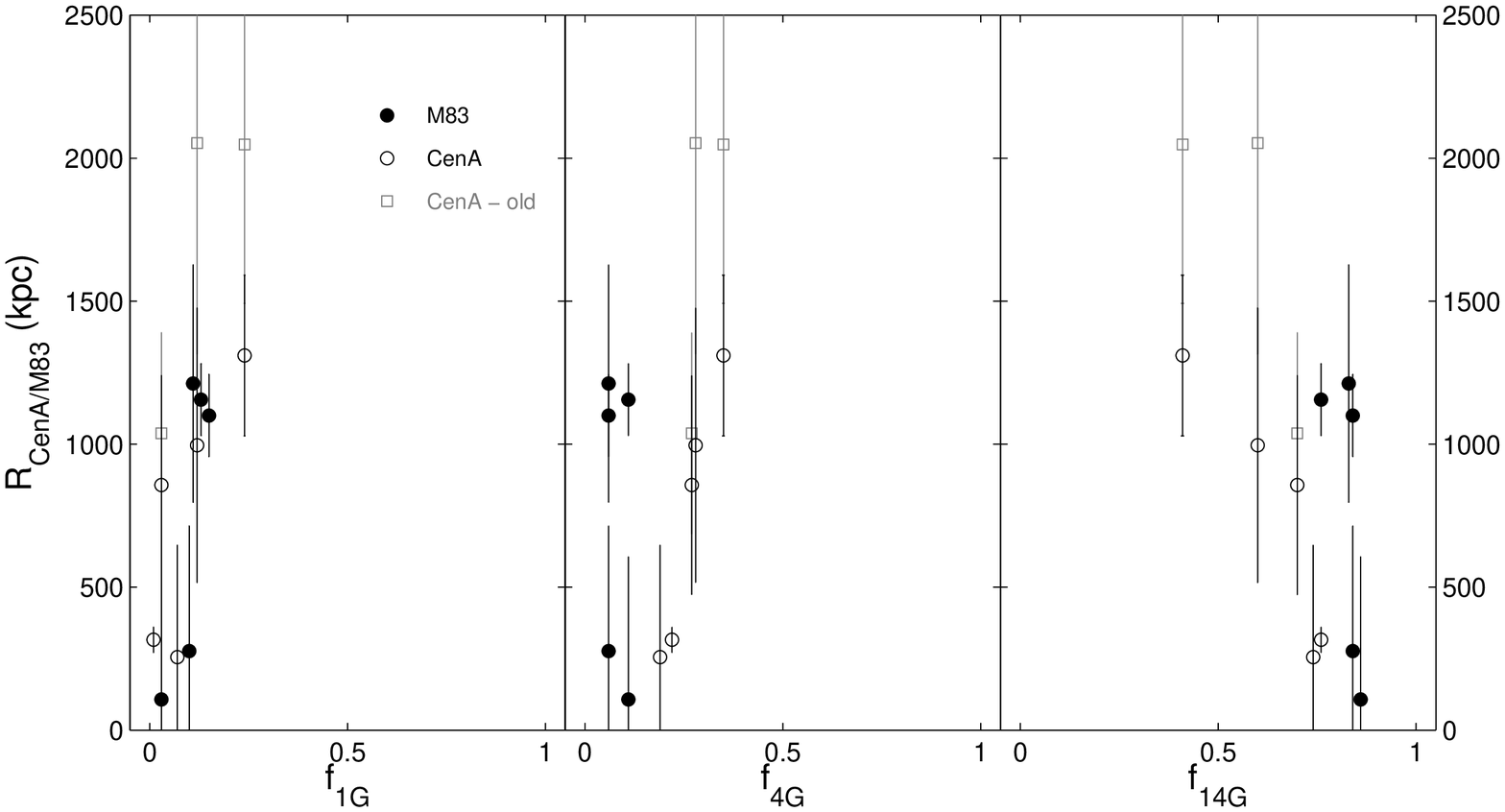}
  \includegraphics[width=11cm]{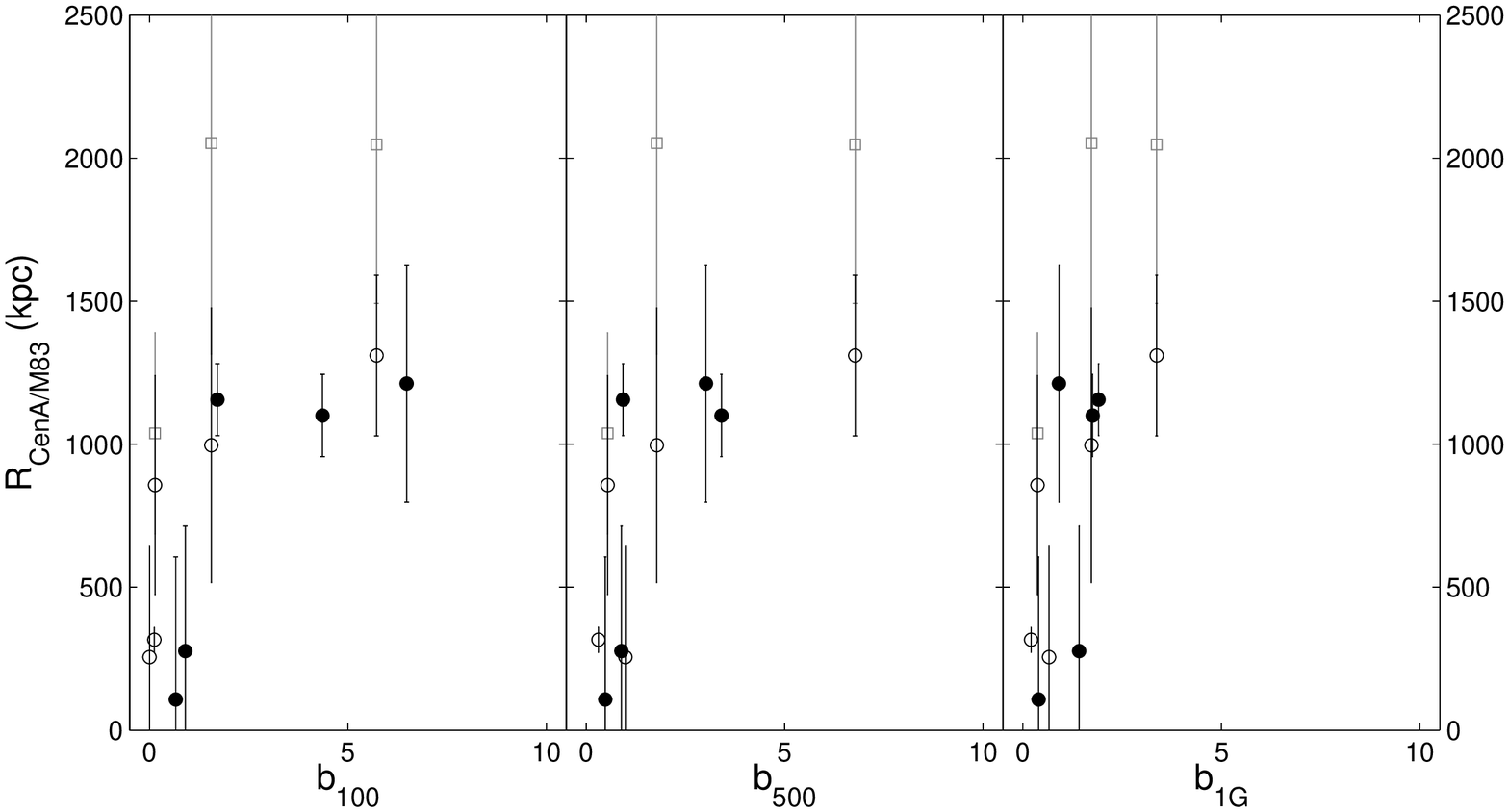}}
\caption{\footnotesize{Deprojected distance from the dominant subgroup galaxy (either CenA or M83) plotted as a function of: (\emph{upper panel}) the fraction of stars formed in the past $0-1$ Gyr ($f_{1G}$), $1-4$ Gyr ($f_{4G}$) and $4-14$ Gyr ($f_{14G}$); (\emph{lower panel}) the ratio of \emph{SFR} to the lifetime average \emph{SFR} over the last 100 Myr ($b_{100}$), 500 Myr ($b_{500}$) and 1 Gyr ($b_{1G}$). The black open circles represent the CenA subgroup late-type dwarfs and the black filled circles are M83 subgroup members. Grey squares indicate the old distances for three of the CenA subgroup members, while black open circles for the same galaxies are new distances that we recompute because they are actually closer to M83 than to CenA (see also Fig.~\ref{sfr_vs_lt}).}} 
\label{dist_vs_lt}
\end{figure*}

We further plot the tidal index as a function of the $f$ and $b$ parameters in Fig.~\ref{tid_vs_lt}. If we do not consider Holmberg IX, we can see that there is a clear trend between the tidal index and both $b_{100}$ and $b_{500}$ (which tells us how active a galaxy has been in the past 100 and 500 Myr, respectively). There is also a hint for such a trend with $b_{1G}$, but less pronounced. The conclusion is that galaxies with a positive tidal index have generally had a low activity in their most recent SFHs, while the ones that are more isolated have a range of different properties. The result confirms what has already been found by \citet{bouchard08} about the current \emph{SFR} being lower in denser environments. They use a large sample of dwarfs in the Local Group, the CenA group and the Sculptor group, relating the dwarfs' physical properties to the local luminosity density. The ANGST dataset, on the other hand, comprises the CMDs of $\sim60$ dwarf galaxies within $\sim4$ Mpc, thus extending the \citet{weisz08} results that we overplot in our Fig.~\ref{tid_vs_lt}. In \citet{weisz11} the SFHs for these objects are presented, but the $b$ parameters for ages $<1$ Gyr that we use for our plots are not yet analyzed (see Weisz et al., in prep.). From a quantitative comparison with the most recent ($\sim1$ Gyr) bins of the SFHs in the ANGST dIrrs, we can however already see that the \emph{SFR}s within this timescale are not inconsistent with their average lifetime value, thus confirming the lack of an enhanced \emph{SFR} for objects with a positive tidal index.

Considering the fraction of stars born at different ages (lower panel of Fig.~\ref{tid_vs_lt}), the $f_{4G}$ and $f_{14G}$ parameters present a larger scatter for a tidal index $\lesssim0.5$. This translates into a common property for galaxies embedded in a denser environment, namely that they have all formed most of their stellar content more than 4 Gyr ago. On the other hand, galaxies that are currently located in less dense environments have a larger range of values for $f_{4G}$ and $f_{14G}$. 

As a last step we also plot the deprojected distance from the dominant group galaxy as a function of the $f$ and $b$ parameters in Fig.~\ref{dist_vs_lt}. For this quantity, only data for the CenA/M83 group are shown. We can see that the results are similar to the ones found for the tidal index. All the dwarf galaxies that are located closer than $\sim\!500$ kpc from a giant galaxy clearly produced stars at a rate much lower than the lifetime average \emph{SFR} in the last $\sim\!1$ Gyr. Their more distant companions, on the other hand, produced stars with a range of different rates. This effect is particularly pronounced within the last 100 Myr. As before, this still holds when considering the fraction of stars born at different times. 

We conclude that the star formation in late-type dwarfs has a considerable range of properties in the field, while those objects that live in an environment with a high galaxy density and close to a giant galaxy had both a lower neutral gas mass (per unit baryonic mass, see previous subsection) and a lower \emph{SFR} with respect to the lifetime average within the last $\sim0.5$--1 Gyr. The details of the SFHs will of course depend on the details of the galactic orbit within the group, but the orbits of our targets are unknown due to their distances. \citet{bouchard08} suggest that the proximity to a large number of galaxies (i.e., a dense group environment) could be accompanied by a sufficiently dense intergalactic medium in order for ram-pressure stripping to happen, and in the CenA/M83 group this could be a possible culprit for the observed suppression of star formation. According to the same authors, another possibility is that galaxies in a dense environment have a higher fraction of ionized gas, because of a higher ionizing flux coming from the neighboring galaxies.


\section{Conclusions} \label{conclus}

In the present study we derive the physical properties for five dwarf irregular galaxies companions of the giant elliptical CenA. Together with results for another sample of five dwarf companions of the giant spiral M83 \citep[presented in][]{crnojevic11b}, we compare the properties derived in galaxies of the CenA/M83 group to the ones found in different environments.

We have analyzed archival HST/ACS data in order to study the resolved stellar populations of these dwarfs. We derive their star formation histories via the synthetic color-magnitude diagram modeling technique adopting the Padova stellar evolutionary models \citep{marigo08}. Our results cofirm the general trend expected for this type of objects, where a low and constantly present star formation activity is alternating with episodes of enhanced activity, with intensities of up to several times the average lifetime value and lasting up to a few $10^2$ Myr (as seen within the past $\sim1$Gyr, where our age resolution is highest). Individually, our resulting star formation histories all show differing shapes, with a range of average star formation rates (from $\sim10^{-3}$ up to $\sim7\times10^{-2}$M$_\odot$ yr$^{-1}$) and low metallicities ([Fe/H]$\sim-1.4$ to $\sim-1.0$). For the target galaxies a fraction between $35\%$ and $60\%$ of the total stellar population was born at times larger than $\sim5$ Gyr ago. Finally, for each dwarf we produce density maps for subsamples of the stellar population with different ages, in order to study how star formation proceeds in these objects. The youngest populations are found in pockets of actively star forming regions located close to the center of the galaxy, while the old stars had time to migrate and redistribute over the whole galaxy's extent assuming a more regular distribution.

We then consider the total sample of ten galaxies, i.e. both M83 and CenA dwarf companions, which comprise four galaxies in dense regions of the group and six rather isolated objects in its outskirts. Our goal is to look for correlations between the parameters characterizing the star formation efficiency at different epochs and other physical properties of the target galaxies. We check whether the \emph{average} star formation rate correlates with galaxy baryonic mass, degree of isolation and deprojected distance from the dominant galaxy. We find it to be higher for higher baryonic mass objects, thus supporting the role of intrinsic properties in the evolution of dwarfs. On the other hand, there is no clear trend of \emph{average} star formation rate with the position of the galaxies in the group. This is a reasonable result, given that the average properties depend on the whole galaxy's history, and we only have information about the current position of the objects within the group because their orbits are unknown. 

We further investigate the dependence on environment of parameters characterizing the star formation efficiency and the fractions of stars produced at different epochs. We intriguingly see that the ratio of the \emph{recent} star formation (last $\sim\!0.5-1$ Gyr) to the lifetime average value does correlate with environment, confirming previous studies. Namely, dwarfs that are closest to the dominant giant galaxy and found in a denser environment have lower such ratios compared to dwarfs in the outskirts. We can thus conclude that the star formation within the last $\sim$1 Gyr is reduced in dense group environments, while more isolated objects show a vast range of different star formation histories (i.e., their evolution is primarily regulated by intrinsic properties). This result is complemented by data for similar luminosity dwarfs in the Local Group and M81 group, which reinforces our conclusions. We find no significant differences among the three groups of galaxies when looking at the galaxies' individual star formation efficiencies at various time epochs, and also looking at the shape of the star formation histories themselves.

Finally, we confirm that both the ratio of neutral gas mass to baryonic mass and the ratio of neutral gas mass to average star formation rate (i.e., the current consumption rate of gas) in the CenA/M83 group are significantly lower for galaxies close to a giant companion. This underlines again the important role that a dense environment plays in the evolution of its dwarf inhabitants \citep[see also][]{grebel03}. In summary, it is evident that both nature and nurture heavily contribute in shaping the star formation histories of our targets.


\begin{acknowledgements}

We would like to thank an anonymous referee for helpful and constructive comments that improved the manuscript. DC is thankful to S. Pasetto for helpful conversations and support. This project was carried out with financial support from the Max Planck Institute for Astronomy, as part of the International Max Planck Research Student program; from the Astronomisches Rechen-Institut at Heidelberg University; and from an STFC Rolling Grant. DC acknowledges travel support from the Heidelberg Graduate School for Fundamental Physics of the University of Heidelberg. EKG acknowledges partial support by the Sonderforschungsbereich ``The Milky Way System'' (SFB881) of the German Research Foundation (DFG) at Heidelberg University (esp. subproject A2). This work is based on observations made with the NASA/ESA Hubble Space Telescope, obtained from the data archive at the Space Telescope Science Institute. STScI is operated by the Association of Universities for Research in Astronomy, Inc. under NASA contract NAS 5-26555. This research made use of the NASA/IPAC Extragalactic Database (NED), which is operated by the Jet Propulsion Laboratory, California Institute of Technology, under contract with the National Aeronautics and Space Administration.

\end{acknowledgements}


\bibliographystyle{aa}
\bibliography{biblio.bib}

\begin{thebibliography}{74}
\expandafter\ifx\csname natexlab\endcsname\relax\def\natexlab#1{#1}\fi

\bibitem[{{Aparicio} \& {Hidalgo}(2009)}]{aparicio09}
{Aparicio}, A. \& {Hidalgo}, S.~L. 2009, \aj, 138, 558

\bibitem[{{Banks} {et~al.}(1999){Banks}, {Disney}, {Knezek}, {Jerjen},
  {Barnes}, {Bhatal}, {de Blok}, {Boyce}, \& {et al.}}]{banks99}
{Banks}, G.~D., {Disney}, M.~J., {Knezek}, P.~M., {et~al.} 1999, \apj, 524, 612

\bibitem[{{Bastian} {et~al.}(2011){Bastian}, {Weisz}, {Skillman}, {McQuinn},
  {Dolphin}, {Gutermuth}, {Cannon}, {Ercolano}, {Gieles}, {Kennicutt}, \&
  {Walter}}]{bastian11}
{Bastian}, N., {Weisz}, D.~R., {Skillman}, E.~D., {et~al.} 2011, \mnras, 412,
  1539

\bibitem[{{Battinelli} {et~al.}(2007){Battinelli}, {Demers}, \&
  {Artigau}}]{battinelli07}
{Battinelli}, P., {Demers}, S., \& {Artigau}, {\'E}. 2007, \aap, 466, 875

\bibitem[{{Beaulieu} {et~al.}(2006){Beaulieu}, {Freeman}, {Carignan},
  {Lockman}, \& {Jerjen}}]{beaulieu06}
{Beaulieu}, S.~F., {Freeman}, K.~C., {Carignan}, C., {Lockman}, F.~J., \&
  {Jerjen}, H. 2006, \aj, 131, 325

\bibitem[{{Bouchard} {et~al.}(2009){Bouchard}, {Da Costa}, \&
  {Jerjen}}]{bouchard08}
{Bouchard}, A., {Da Costa}, G.~S., \& {Jerjen}, H. 2009, \aj, 137, 3038

\bibitem[{{Bouchard} {et~al.}(2007){Bouchard}, {Jerjen}, {Da Costa}, \&
  {Ott}}]{bouchard07}
{Bouchard}, A., {Jerjen}, H., {Da Costa}, G.~S., \& {Ott}, J. 2007, \aj, 133,
  261

\bibitem[{{Chabrier}(2003)}]{chabrier03}
{Chabrier}, G. 2003, \apjl, 586, L133

\bibitem[{{Cignoni} \& {Tosi}(2010)}]{cignoni10}
{Cignoni}, M. \& {Tosi}, M. 2010, Advances in Astronomy, 2010, 3

\bibitem[{{Cole} {et~al.}(2007){Cole}, {Skillman}, {Tolstoy}, {Gallagher},
  {Aparicio}, {Dolphin}, {Gallart}, {Hidalgo}, \& {et al.}}]{cole07}
{Cole}, A.~A., {Skillman}, E.~D., {Tolstoy}, E., {et~al.} 2007, \apjl, 659, L17

\bibitem[{{C{\^o}t{\'e}} {et~al.}(2000){C{\^o}t{\'e}}, {Carignan}, \&
  {Freeman}}]{cote00}
{C{\^o}t{\'e}}, S., {Carignan}, C., \& {Freeman}, K.~C. 2000, \aj, 120, 3027

\bibitem[{{C{\^o}t{\'e}} {et~al.}(2009){C{\^o}t{\'e}}, {Draginda}, {Skillman},
  \& {Miller}}]{cote09}
{C{\^o}t{\'e}}, S., {Draginda}, A., {Skillman}, E.~D., \& {Miller}, B.~W. 2009,
  \aj, 138, 1037

\bibitem[{{Crnojevi{\'c}} {et~al.}(2011{\natexlab{a}}){Crnojevi{\'c}},
  {Grebel}, \& {Cole}}]{crnojevic11b}
{Crnojevi{\'c}}, D., {Grebel}, E.~K., \& {Cole}, A.~A. 2011{\natexlab{a}},
  \aap, 530, A59

\bibitem[{{Crnojevi{\'c}} {et~al.}(2010){Crnojevi{\'c}}, {Grebel}, \&
  {Koch}}]{crnojevic10}
{Crnojevi{\'c}}, D., {Grebel}, E.~K., \& {Koch}, A. 2010, \aap, 516, A85

\bibitem[{{Crnojevi{\'c}} {et~al.}(2011{\natexlab{b}}){Crnojevi{\'c}},
  {Rejkuba}, {Grebel}, {da Costa}, \& {Jerjen}}]{crnojevic11a}
{Crnojevi{\'c}}, D., {Rejkuba}, M., {Grebel}, E.~K., {da Costa}, G., \&
  {Jerjen}, H. 2011{\natexlab{b}}, \aap, 530, A58

\bibitem[{{Dalcanton} {et~al.}(2009){Dalcanton}, {Williams}, {Seth}, {Dolphin},
  {Holtzman}, {Rosema}, {Skillman}, {Cole}, \& {et al.}}]{dalcanton09}
{Dalcanton}, J.~J., {Williams}, B.~F., {Seth}, A.~C., {et~al.} 2009, \apjs,
  183, 67

\bibitem[{{Davidge}(2007)}]{davidge07}
{Davidge}, T.~J. 2007, \aj, 134, 1799

\bibitem[{{Dohm-Palmer} {et~al.}(2002){Dohm-Palmer}, {Skillman}, {Mateo},
  {Saha}, {Dolphin}, {Tolstoy}, {Gallagher}, \& {Cole}}]{dohm02}
{Dohm-Palmer}, R.~C., {Skillman}, E.~D., {Mateo}, M., {et~al.} 2002, \aj, 123,
  813

\bibitem[{{Dohm-Palmer} {et~al.}(1997){Dohm-Palmer}, {Skillman}, {Saha},
  {Tolstoy}, {Mateo}, {Gallagher}, {Hoessel}, {Chiosi}, \& {et al.}}]{dohm97}
{Dohm-Palmer}, R.~C., {Skillman}, E.~D., {Saha}, A., {et~al.} 1997, \aj, 114,
  2527

\bibitem[{{Dolphin}(2002)}]{dolphin02}
{Dolphin}, A.~E. 2002, \mnras, 332, 91

\bibitem[{{Dotter} {et~al.}(2008){Dotter}, {Chaboyer}, {Jevremovi{\'c}},
  {Kostov}, {Baron}, \& {Ferguson}}]{dotter08}
{Dotter}, A., {Chaboyer}, B., {Jevremovi{\'c}}, D., {et~al.} 2008, \apjs, 178,
  89

\bibitem[{{Duc} {et~al.}(2000){Duc}, {Brinks}, {Springel}, {Pichardo},
  {Weilbacher}, \& {Mirabel}}]{duc00}
{Duc}, P., {Brinks}, E., {Springel}, V., {et~al.} 2000, \aj, 120, 1238

\bibitem[{{Duc} {et~al.}(1997){Duc}, {Brinks}, {Wink}, \& {Mirabel}}]{duc97}
{Duc}, P., {Brinks}, E., {Wink}, J.~E., \& {Mirabel}, I.~F. 1997, \aap, 326,
  537

\bibitem[{{Duquennoy} \& {Mayor}(1991)}]{duque91}
{Duquennoy}, A. \& {Mayor}, M. 1991, \aap, 248, 485

\bibitem[{{Einasto} {et~al.}(1974){Einasto}, {Saar}, {Kaasik}, \&
  {Chernin}}]{einasto74}
{Einasto}, J., {Saar}, E., {Kaasik}, A., \& {Chernin}, A.~D. 1974, \nat, 252,
  111

\bibitem[{{Gallart} {et~al.}(2005){Gallart}, {Zoccali}, \&
  {Aparicio}}]{gallart05}
{Gallart}, C., {Zoccali}, M., \& {Aparicio}, A. 2005, \araa, 43, 387

\bibitem[{{Georgiev} {et~al.}(2008){Georgiev}, {Goudfrooij}, {Puzia}, \&
  {Hilker}}]{georgiev08}
{Georgiev}, I.~Y., {Goudfrooij}, P., {Puzia}, T.~H., \& {Hilker}, M. 2008, \aj,
  135, 1858

\bibitem[{{Girardi} {et~al.}(2005){Girardi}, {Groenewegen}, {Hatziminaoglou},
  \& {da Costa}}]{girardi05}
{Girardi}, L., {Groenewegen}, M.~A.~T., {Hatziminaoglou}, E., \& {da Costa}, L.
  2005, \aap, 436, 895

\bibitem[{{Glatt} {et~al.}(2010){Glatt}, {Grebel}, \& {Koch}}]{glatt10}
{Glatt}, K., {Grebel}, E.~K., \& {Koch}, A. 2010, \aap, 517, A50

\bibitem[{{Glatt} {et~al.}(2008){Glatt}, {Grebel}, {Sabbi}, {Gallagher},
  {Nota}, {Sirianni}, {Clementini}, {Tosi}, \& {et al.}}]{glatt08b}
{Glatt}, K., {Grebel}, E.~K., {Sabbi}, E., {et~al.} 2008, \aj, 136, 1703

\bibitem[{{Grebel}(2000)}]{grebel00esa}
{Grebel}, E.~K. 2000, in ESA Special Publication, Vol. 445, Star Formation from
  the Small to the Large Scale, ed. {F.~Favata, A.~Kaas, \& A.~Wilson}, 87

\bibitem[{{Grebel}(2001)}]{grebel01}
{Grebel}, E.~K. 2001, Astrophysics and Space Science Supplement, 277, 231

\bibitem[{{Grebel}(2004)}]{grebel04a}
{Grebel}, E.~K. 2004, in Origin and Evolution of the Elements, Carnegie
  Observatories Centennial Symposia, Cambridge University Press, ed.
  {A.~McWilliam \& M.~Rauch}, 234

\bibitem[{{Grebel} \& {Brandner}(1998)}]{grebel98}
{Grebel}, E.~K. \& {Brandner}, W. 1998, in Magellanic Clouds and Other Dwarf
  Galaxies, ed. {T.~Richtler \& J.~M.~Braun, Shaker Verlag}, 151

\bibitem[{{Grebel} \& {Gallagher}(2004)}]{grebel04}
{Grebel}, E.~K. \& {Gallagher}, III, J.~S. 2004, \apjl, 610, L89

\bibitem[{{Grebel} {et~al.}(2003){Grebel}, {Gallagher}, \&
  {Harbeck}}]{grebel03}
{Grebel}, E.~K., {Gallagher}, III, J.~S., \& {Harbeck}, D. 2003, \aj, 125, 1926

\bibitem[{{Harbeck} {et~al.}(2001){Harbeck}, {Grebel}, {Holtzman},
  {Guhathakurta}, {Brandner}, {Geisler}, {Sarajedini}, {Dolphin}, \& {et
  al.}}]{harbeck01}
{Harbeck}, D., {Grebel}, E.~K., {Holtzman}, J., {et~al.} 2001, \aj, 122, 3092

\bibitem[{{Hidalgo} {et~al.}(2011){Hidalgo}, {Aparicio}, {Skillman}, {Monelli},
  {Gallart}, {Cole}, {Dolphin}, {Weisz}, {Bernard}, {Cassisi}, {Mayer},
  {Stetson}, {Tolstoy}, \& {Ferguson}}]{hidalgo11}
{Hidalgo}, S.~L., {Aparicio}, A., {Skillman}, E., {et~al.} 2011, \apj, 730, 14

\bibitem[{{Israel}(1998)}]{israel98}
{Israel}, F.~P. 1998, \aapr, 8, 237

\bibitem[{{Jerjen} {et~al.}(2000){Jerjen}, {Binggeli}, \&
  {Freeman}}]{jerjen00b}
{Jerjen}, H., {Binggeli}, B., \& {Freeman}, K.~C. 2000, \aj, 119, 593

\bibitem[{{Karachentsev}(2005)}]{kara05}
{Karachentsev}, I.~D. 2005, \aj, 129, 178

\bibitem[{{Karachentsev} {et~al.}(2002{\natexlab{a}}){Karachentsev}, {Dolphin},
  {Geisler}, {Grebel}, {Guhathakurta}, {Hodge}, {Karachentseva}, {Sarajedini},
  \& {et al.}}]{kara02_m81}
{Karachentsev}, I.~D., {Dolphin}, A.~E., {Geisler}, D., {et~al.}
  2002{\natexlab{a}}, \aap, 383, 125

\bibitem[{{Karachentsev} {et~al.}(2003{\natexlab{a}}){Karachentsev}, {Grebel},
  {Sharina}, {Dolphin}, {Geisler}, {Guhathakurta}, {Hodge}, {Karachentseva}, \&
  {et al.}}]{kara03_sc}
{Karachentsev}, I.~D., {Grebel}, E.~K., {Sharina}, M.~E., {et~al.}
  2003{\natexlab{a}}, \aap, 404, 93

\bibitem[{{Karachentsev} {et~al.}(2004){Karachentsev}, {Karachentseva},
  {Huchtmeier}, \& {Makarov}}]{kara04}
{Karachentsev}, I.~D., {Karachentseva}, V.~E., {Huchtmeier}, W.~K., \&
  {Makarov}, D.~I. 2004, \aj, 127, 2031

\bibitem[{{Karachentsev} {et~al.}(2002{\natexlab{b}}){Karachentsev}, {Sharina},
  {Dolphin}, {Grebel}, {Geisler}, {Guhathakurta}, {Hodge}, {Karachentseva}, \&
  {et al.}}]{kara02}
{Karachentsev}, I.~D., {Sharina}, M.~E., {Dolphin}, A.~E., {et~al.}
  2002{\natexlab{b}}, \aap, 385, 21

\bibitem[{{Karachentsev} {et~al.}(2003{\natexlab{b}}){Karachentsev}, {Sharina},
  {Dolphin}, {Grebel}, {Geisler}, {Guhathakurta}, {Hodge}, {Karachentseva}, \&
  {et al.}}]{kara03_can}
{Karachentsev}, I.~D., {Sharina}, M.~E., {Dolphin}, A.~E., {et~al.}
  2003{\natexlab{b}}, \aap, 398, 467

\bibitem[{{Karachentsev} {et~al.}(2007){Karachentsev}, {Tully}, {Dolphin},
  {Sharina}, {Makarova}, {Makarov}, {Sakai}, {Shaya}, \& {et al.}}]{kara07}
{Karachentsev}, I.~D., {Tully}, R.~B., {Dolphin}, A., {et~al.} 2007, \aj, 133,
  504

\bibitem[{{Kazantzidis} {et~al.}(2011){Kazantzidis}, {{\L}okas}, {Callegari},
  {Mayer}, \& {Moustakas}}]{kazantzidis11}
{Kazantzidis}, S., {{\L}okas}, E.~L., {Callegari}, S., {Mayer}, L., \&
  {Moustakas}, L.~A. 2011, \apj, 726, 98

\bibitem[{{Koleva} {et~al.}(2009){Koleva}, {de Rijcke}, {Prugniel},
  {Zeilinger}, \& {Michielsen}}]{koleva09}
{Koleva}, M., {de Rijcke}, S., {Prugniel}, P., {Zeilinger}, W.~W., \&
  {Michielsen}, D. 2009, \mnras, 396, 2133

\bibitem[{{Lee} {et~al.}(2007){Lee}, {Zucker}, \& {Grebel}}]{lee07}
{Lee}, H., {Zucker}, D.~B., \& {Grebel}, E.~K. 2007, \mnras, 376, 820

\bibitem[{{Lianou} {et~al.}(2010){Lianou}, {Grebel}, \& {Koch}}]{lianou10}
{Lianou}, S., {Grebel}, E.~K., \& {Koch}, A. 2010, \aap, 521, A43

\bibitem[{{Makarova} {et~al.}(2010){Makarova}, {Koleva}, {Makarov}, \&
  {Prugniel}}]{makarova10}
{Makarova}, L., {Koleva}, M., {Makarov}, D., \& {Prugniel}, P. 2010, \mnras,
  406, 1152

\bibitem[{{Makarova} {et~al.}(2002){Makarova}, {Karachentsev}, {Grebel}, \&
  {Barsunova}}]{makarova02}
{Makarova}, L.~N., {Karachentsev}, I.~D., {Grebel}, E.~K., \& {Barsunova},
  O.~Y. 2002, \aap, 384, 72

\bibitem[{{Marconi} {et~al.}(1995){Marconi}, {Tosi}, {Greggio}, \&
  {Focardi}}]{marconi95}
{Marconi}, G., {Tosi}, M., {Greggio}, L., \& {Focardi}, P. 1995, \aj, 109, 173

\bibitem[{{Marigo} {et~al.}(2008){Marigo}, {Girardi}, {Bressan}, {Groenewegen},
  {Silva}, \& {Granato}}]{marigo08}
{Marigo}, P., {Girardi}, L., {Bressan}, A., {et~al.} 2008, \aap, 482, 883

\bibitem[{{Matthews} {et~al.}(1998){Matthews}, {van Driel}, \&
  {Gallagher}}]{matthews98}
{Matthews}, L.~D., {van Driel}, W., \& {Gallagher}, III, J.~S. 1998, \aj, 116,
  2196

\bibitem[{{Mayer} {et~al.}(2006){Mayer}, {Mastropietro}, {Wadsley}, {Stadel},
  \& {Moore}}]{mayer06}
{Mayer}, L., {Mastropietro}, C., {Wadsley}, J., {Stadel}, J., \& {Moore}, B.
  2006, \mnras, 369, 1021

\bibitem[{{Mazeh} {et~al.}(1992){Mazeh}, {Goldberg}, {Duquennoy}, \&
  {Mayor}}]{mazeh92}
{Mazeh}, T., {Goldberg}, D., {Duquennoy}, A., \& {Mayor}, M. 1992, \apj, 401,
  265

\bibitem[{{McQuinn} {et~al.}(2010){McQuinn}, {Skillman}, {Cannon}, {Dalcanton},
  {Dolphin}, {Hidalgo-Rodr{\'{\i}}guez}, {Holtzman}, {Stark}, {Weisz}, \&
  {Williams}}]{mcquinn10}
{McQuinn}, K.~B.~W., {Skillman}, E.~D., {Cannon}, J.~M., {et~al.} 2010, \apj,
  721, 297

\bibitem[{{McQuinn} {et~al.}(2009){McQuinn}, {Skillman}, {Cannon}, {Dalcanton},
  {Dolphin}, {Stark}, \& {Weisz}}]{mcquinn09}
{McQuinn}, K.~B.~W., {Skillman}, E.~D., {Cannon}, J.~M., {et~al.} 2009, \apj,
  695, 561

\bibitem[{{McQuinn} {et~al.}(2011){McQuinn}, {Skillman}, {Dalcanton},
  {Dolphin}, {Holtzman}, {Weisz}, \& {Williams}}]{mcquinn11}
{McQuinn}, K.~B.~W., {Skillman}, E.~D., {Dalcanton}, J.~J., {et~al.} 2011,
  \apj, 740, 48

\bibitem[{{Pasetto} {et~al.}(2011){Pasetto}, {Grebel}, {Berczik}, {Chiosi}, \&
  {Spurzem}}]{pasetto11}
{Pasetto}, S., {Grebel}, E.~K., {Berczik}, P., {Chiosi}, C., \& {Spurzem}, R.
  2011, \aap, 525, A99

\bibitem[{{Phillips} {et~al.}(1986){Phillips}, {Jenkins}, {Dopita}, {Sadler},
  \& {Binette}}]{phillips86}
{Phillips}, M.~M., {Jenkins}, C.~R., {Dopita}, M.~A., {Sadler}, E.~M., \&
  {Binette}, L. 1986, \aj, 91, 1062

\bibitem[{{Schlegel} {et~al.}(1998){Schlegel}, {Finkbeiner}, \&
  {Davis}}]{schlegel98}
{Schlegel}, D.~J., {Finkbeiner}, D.~P., \& {Davis}, M. 1998, \apj, 500, 525

\bibitem[{{Seiden} {et~al.}(1979){Seiden}, {Schulman}, \& {Gerola}}]{seiden79}
{Seiden}, P.~E., {Schulman}, L.~S., \& {Gerola}, H. 1979, \apj, 232, 702

\bibitem[{{Sharina} {et~al.}(2008){Sharina}, {Karachentsev}, {Dolphin},
  {Karachentseva}, {Tully}, {Karataeva}, {Makarov}, {Makarova}, \& {et
  al.}}]{sharina08}
{Sharina}, M.~E., {Karachentsev}, I.~D., {Dolphin}, A.~E., {et~al.} 2008,
  \mnras, 384, 1544

\bibitem[{{Skillman} {et~al.}(2003){Skillman}, {Tolstoy}, {Cole}, {Dolphin},
  {Saha}, {Gallagher}, {Dohm-Palmer}, \& {Mateo}}]{skillman03}
{Skillman}, E.~D., {Tolstoy}, E., {Cole}, A.~A., {et~al.} 2003, \apj, 596, 253

\bibitem[{{Tolstoy} {et~al.}(2009){Tolstoy}, {Hill}, \& {Tosi}}]{tolstoy09}
{Tolstoy}, E., {Hill}, V., \& {Tosi}, M. 2009, \araa, 47, 371

\bibitem[{{Toomre} \& {Toomre}(1972)}]{toomre72}
{Toomre}, A. \& {Toomre}, J. 1972, \apj, 178, 623

\bibitem[{{Trentham} \& {Tully}(2002)}]{trentham02}
{Trentham}, N. \& {Tully}, R.~B. 2002, \mnras, 335, 712

\bibitem[{{van Dyk} {et~al.}(1998){van Dyk}, {Puche}, \& {Wong}}]{vandyk98}
{van Dyk}, S.~D., {Puche}, D., \& {Wong}, T. 1998, \aj, 116, 2341

\bibitem[{{Weisz} {et~al.}(2011){Weisz}, {Dalcanton}, {Williams}, {Gilbert},
  {Skillman}, {Seth}, {Dolphin}, {McQuinn}, {Gogarten}, {Holtzman}, {Rosema},
  {Cole}, {Karachentsev}, \& {Zaritsky}}]{weisz11}
{Weisz}, D.~R., {Dalcanton}, J.~J., {Williams}, B.~F., {et~al.} 2011, \apj,
  739, 5

\bibitem[{{Weisz} {et~al.}(2008){Weisz}, {Skillman}, {Cannon}, {Dolphin},
  {Kennicutt}, {Lee}, \& {Walter}}]{weisz08}
{Weisz}, D.~R., {Skillman}, E.~D., {Cannon}, J.~M., {et~al.} 2008, \apj, 689,
  160

\bibitem[{{Zaritsky} {et~al.}(2000){Zaritsky}, {Harris}, {Grebel}, \&
  {Thompson}}]{zaritsky00}
{Zaritsky}, D., {Harris}, J., {Grebel}, E.~K., \& {Thompson}, I.~B. 2000,
  \apjl, 534, L53

\end{thebibliography}


\end{document}